\documentclass[11pt,a4paper]{article}
\pdfoutput=1
\usepackage{jheppub}
\usepackage{hyperref}
\usepackage{amssymb}
\usepackage{amsmath}
\usepackage{latexsym}

\newcommand{\de}{\partial} 
\newcommand{\ket}[1]{\lvert #1 \rangle} 
\newcommand{\bra}[1]{\langle #1 \rvert} 
\newcommand{\puntos}[1]{^{\bullet} \!\! #1} 
\newcommand{\puntod}[1]{ #1^{\!\! \bullet} } 
\newcommand{\puntods}[1]{^{\bullet} \!\! #1^{\!\! \bullet} } 

\newcommand{\media}[1]{\langle #1 \rangle} 
\newcommand{\Dc}{\mathcal{D}}
\newcommand{\A}{\mathcal{A}}
\newcommand{\Tr}{\mathrm{Tr}}
\newcommand{\be}{\begin{equation}}
\newcommand{\ee}{\end{equation}}
\newcommand{\bea}{\begin{eqnarray}}
\newcommand{\eea}{\end{eqnarray}}

\def\bec{\begin{center}}
\def\ec{\end{center}}
\def\a{\alpha}  

\def\b{\beta}   
 
\def\C{\Gamma}
\def\d{\delta}

\def\l{\lambda}
\def\L{\Lambda}
\def\m{\mu}
\def\n{\nu}
\def\r{\rho}
\def\s{\sigma}

\def\t{\tau}

\def\del{\partial}

\def\nn{\nonumber}
\newcommand{\eq}[1]{(\ref{#1})}

\title{Quantum theories of $\bf (p,q)$-forms}
\author{Fiorenzo Bastianelli,}
\author{Roberto Bonezzi}
\author{and Carlo Iazeolla}
\affiliation{Dipartimento  di Fisica, Universit{\`a} di Bologna and\\
INFN, Sezione di Bologna, via Irnerio 46, I-40126 Bologna, Italy}

\emailAdd{bastianelli@bo.infn.it}\emailAdd{bonezzi@bo.infn.it}\emailAdd{iazeolla@bo.infn.it}

\abstract{We describe quantum theories for massless $(p,q)$-forms living on K\"ahler spaces. In particular we consider four different types of quantum theories:
two types involve gauge symmetries and two types are simpler theories without  gauge invariances. The latter can be seen as building blocks of the former.
Their equations of motion can be obtained in a natural way by first-quantizing a spinning particle with a U(2)-extended supersymmetry on the worldline.
The particle system contains four supersymmetric charges, represented quantum mechanically by the  Dolbeault operators $\partial$, $\bar \partial$,
and their hermitian conjugates $\partial^\dagger$, $\bar \partial^\dagger$. After studying how the $(p,q)$-form field theories emerge from the particle system,
we investigate their one loop effective actions, identify corresponding heat kernel coefficients, and derive exact duality relations.
The dualities are seen to include mismatches related to topological indices  and analytic torsions, which are computed as
${\rm Tr}\,(-1)^F$ and ${\rm Tr}\, (-1)^F F$ in the first quantized  supersymmetric nonlinear sigma model for a suitable fermion number operator $F$.}


\begin{document}
\maketitle
\flushbottom

\section{Introduction}

The aim of this paper is to investigate several quantum theories for $(p,q)$-forms living on K\"ahler spaces and study their duality relations at the quantum level.
The most interesting ones contain gauge invariances and generalize in a natural way the theory of differential $p$-forms  $C$ with
equations of motion  of the Maxwell type $d^\dagger d C=0$
and gauge invariance $\delta C = d \Lambda$.
For an arbitrary  $(p,q)$-form $A$ they read
\be \label{intro-A}
\left(\de^\dagger\de -\bar\de\bar\de^\dagger\right)A=\de\bar\de\rho
\ee
where $\partial$ and $\bar \partial$ are the   Dolbeault operators with hermitian conjugate $\partial^\dagger$ and $\bar \partial^\dagger$,
and $\rho$ is a compensator for the gauge symmetry.  For  the particular case of
a $(p,d-2-p)$-form $\alpha$,
one can eliminate the need of a compensator by suitably modifying the equations to
\be \label{intro-alpha}
\left(\de^\dagger\de -\bar\de\bar\de^\dagger+\de\bar\de\Tr\right)\alpha=0
\ee
where ``Tr" takes a trace of the form (with a suitable normalization to be defined later).
Simpler theories without gauge invariances can be defined in terms of the Laplace-Beltrami operator  $\triangle$
on $(p,q)$-forms $B$
\be
\triangle B=0
\ee
and can be seen as building blocks for the quantum theories of the former.

These theories can be obtained from the quantization of the U(2) spinning particle, originally presented in
\cite{Marcus:1994em},  extended to the U($N$) case in  \cite{Marcus:1994mm},
and analyzed in \cite{Bastianelli:2009vj} to identify a class of higher spin equations on complex manifolds (see also \cite{Iazeolla:2007wt} for a related work on complex interacting higher spin field equations).
For $N\leq 2$ these higher spin equations reduce to equations for ordinary differential $(p,q)$-forms on complex manifolds with K\"ahler metrics.
In particular, the case $N=1$  describes $(p,0)$-forms
and has been discussed in \cite{Bastianelli:2011pe}.
The ensuing equations have the property that one may allow for an arbitrary coupling to the U(1) part of the $U(d)$ holonomy group of the K\"ahler space.
For a special value of this coupling the set of all the $(p,0)$-forms is equivalent to a fermionic Dirac field.
In the present paper we wish to develop a similar analysis for the case $N=2$, which is rich enough to provide a description of  arbitrary $(p,q)$-forms.
An analogous treatment for standard $p$-forms on real manifolds was given in
\cite{Bastianelli:2005vk} and \cite{Bastianelli:2005uy}, where the worldline techniques
of  \cite{Bastianelli:2002fv, Bastianelli:2002qw,Bastianelli:2003bg}
were used to study the one loop effective actions as function of the background metric. Such a treatment provides a useful
guideline for the present analysis as well.

The gauge invariant quantum theories of $(p,q)$-forms considered here contain
some structural elements present also in the theory of higher spin fields,
such as the appearance of compensators to maintain unconstrained gauge invariance \cite{Francia:2002aa,Francia:2002pt},
as in eq. (\ref{intro-A}), or the emergence of  Fronsdal-like equations of motion \cite{Fronsdal:1978rb}, as in eq. (\ref{intro-alpha}).
A property of the present equations is that they can be defined on arbitrary K\"ahler spaces. On the contrary,
higher spin equations on complex spaces defined by the U($N$) spinning particle
require for $N>2$ special constraints on the background \cite{Bastianelli:2009vj}.
This is analogous to the standard theory of higher spin equations described by  the O($N$) spinning particle,
which for $N>2$ requires a  conformally flat spacetime \cite{Bastianelli:2008nm}, including  (A)dS as particular cases \cite{Kuzenko:1995mg}.

Some of the equations presented here, those of the $\alpha$-type as in (\ref{intro-alpha}), were already introduced in \cite{Bastianelli:2009vj}
using a different basis for the fields, namely a multi-form with two set of  $p$ antisymmetric holomorphic indices, rather than a  $(p, d-2-p)$-form.
The basis of $(p,q)$-forms is perhaps more intuitive geometrically, and will be used here.
Similar equations were also analyzed in \cite{Cherney:2009vg} by making use of the ``detour" construction described in
\cite{Bastianelli:2009eh,Cherney:2009mf,Cherney:2009md}. All of these first quantized pictures make use of
supersymmetric quantum mechanics \cite{Witten:1982df,Witten:1982im}.
In particular, one needs a  supersymmetric quantum mechanics defined by a one-dimensional
nonlinear sigma model with  K\"ahler manifolds as target space and four supercharges that give rise to the
Dolbeault operators $\partial$, $\bar \partial$ and their hermitian conjugates $\partial^\dagger$, $\bar \partial^\dagger$.
This model has been used in  \cite{AlvarezGaume:1983at} to give a supersymmetric proof of the index theorem for the Dolbeault complex
(the index of $\bar \de$ on K\"ahler manifolds), and its lagrangian reviewed for example in  \cite{FigueroaO'Farrill:1997ks}.  For our purposes
the supersymmetry algebra has to be gauged suitably to provide the spinning particle actions that implement
the  worldline description of the various $(p, q)$-form field theories.
Canonical quantization produces  the field equations corresponding to
the particular model that is selected by the charges that one wishes to gauge. Path integral quantization
provides then a simple representation of the corresponding one loop effective action,
which we use for computing the first few heat kernel coefficients.
Dual descriptions are  easily identified in this first quantized picture.
They include topological mismatches that are related to topological indices and analytic torsions, which can be obtained
in the supersymmetric quantum mechanics as Witten indices of the form
 ${\rm Tr}\,(-1)^F$ for a suitable fermion number operator $F$ \cite{Witten:1982df}, and indices of the type
 ${\rm Tr}\, (-1)^F F$, originally introduced in \cite{Cecotti:1992qh}
 for two dimensional supersymmetric field theories, respectively.

We start in section 2 with canonical quantization to obtain the equations of motion
 for the $(p, q)$-forms in flat complex space from the spinning particles. The  susy algebra contains U(2) as maximal R-symmetry group.
 We gauge the full U(2) group to obtain what we call the $\alpha$ and $\beta$ models.
 Gauging a $\rm{U}(1) \times \rm{U}(1)$ subgroup produces instead the A and B models, that can be defined for arbitrary $(p, q)$-forms.
 Gauging the supercharges produces field theories with gauge invariances, the A and $\alpha$ models.
  Keeping the supercharges ungauged  gives rise to simpler models without gauge invariances, the B and $\beta$ models.
  The hamiltonian is always gauged to obtain from first quantization
  the standard proper time representation of propagators and effective actions.
In section 3 we extend the previous results to complex spaces with an arbitrary  K\"ahler  metric, treated as a background field.
In section 4 we quantize the particle models on a circle to obtain one-loop effective actions. We  compute the first few heat kernel coefficients
for all of the models discussed previously using a worldline path integral.
In sections 5 we analyze systematically duality relations, and show how a variety of topological mismatches
arise for the A-type models. We end up with our conclusions in section 6, and include three appendices to provide conventions,
expressions for the worldline propagators, and a list of topological quantities entering the duality relations.


\section{Canonical quantization and equations of motion in flat space}\label{sec flat space}

We are aiming at describing four different quantum theories of differential forms by means of appropriate spinning particle models. The ground on which all these models are constructed is the $N=2$ extended complex superalgebra in one dimension.
The different versions are obtained by gauging suitable subgroups of the whole superalgebra.

We begin by considering such spinning particle models in flat complex space $\mathbb{C}^d$ (see Appendix \ref{app:notations} for notations and conventions). The particle is described by complex coordinates $x^\mu(t), \bar{x}^{\bar\mu}(t)$, $\mu = 1,...,d$, and carries additional degrees of freedom associated to their fermionic superpartners $\psi_i^\mu(t), \bar\psi^i_{\mu}(t)$, $i=1,2$, belonging to the (anti-)fundamental representation of the U(2) $R$-symmetry group.
Spacetime indices are lowered and raised with the flat metric $\delta_{\mu\bar\nu}$ and its inverse, so that we will often use $\bar\psi^{i\bar\mu}=\delta^{\nu\bar\mu}\bar\psi^i_{\nu}$ and $\psi_{i\bar\mu}=\delta_{\nu\bar\mu}\psi_i^\nu$. With the above ingredients, the ungauged model is described by the phase-space action
\begin{equation}\label{ungaction}
S \ = \ \int_0^1dt\Big[p_\mu\dot x^\mu+\bar p_{\bar\mu}\dot{\bar x}^{\bar\mu}+i\bar\psi^i_\mu\dot\psi_i^\mu-p_\mu\bar p^{\mu}\Big]\;,
\end{equation}
which encodes the motion of a free particle with a pseudoclassical spin associated to the Grassmann coordinates. The corresponding conserved charges
\begin{equation}\label{charges}
H \ = \ p_\mu\bar p^{\mu} \ , \qquad Q_i \ = \ \psi^\mu_i p_\mu\ , \qquad \bar Q^i \ = \ \bar\psi^{\bar\mu i}\bar p_{\bar\mu}\ , \qquad J_i^j \ = \ \psi^\mu_i \bar\psi_\mu^j
\end{equation}
generate a $U(2)$-extended supersymmetry algebra on the wordline. Indeed, by using the commutation relations that follow from the classical Poisson bracket algebra
\begin{equation}\label{commrel}
[x^\mu,p_\nu] \ = \ i\delta^\mu_\nu \ , \qquad [\bar x^{\bar\mu},\bar p_{\bar\nu}] \ = \ i\delta^{\bar\mu}_{\bar\nu} \ , \qquad \{\psi^\mu_i,\bar\psi^{j}_\nu\} \ = \ \delta^{\mu}_\nu\,\delta_i^j \ ,
\end{equation}
it is easy to check that the above conserved charges satisfy the quantum commutation relations
\begin{eqnarray}\label{U2susyalg}
 \left\{Q_i,\bar Q^{j}\right\} & = & \delta_i^j H \ , \nonumber\\
 \left[J_i^j,Q_k\right] & = & \delta^j_k Q_i \ , \qquad \ \ \  \left[J_i^j,\bar Q^k\right] \ = \ -\delta^k_i \bar Q^j \ , \\
 \left[J_i^j,J^l_k\right] & = & \delta^j_k J^l_i-\delta^l_i J^j_k \ ,\nonumber
\end{eqnarray}
while other independent graded-commutators vanish. Note that ordering ambiguities are only present in $J_i^j$, for which we choose the graded-symmetric ordering \emph{i.e.}
\be
J_i^j \ = \ \frac{1}{2} (\psi^\mu_i \bar\psi_\mu^j -\bar\psi_\mu^j\psi^\mu_i) \ = \ \psi^\mu_i \bar\psi_\mu^j -\frac{d}{2}\d_i^j
\label{ordering}
\ee
that is naturally reproduced by the fermionic path integral of section \ref{sec effective actions}.
Choosing the Hilbert space basis of $(x^\mu,\bar x^{\bar\mu},\psi_1^\mu,\bar\psi^{\bar\mu 2})$-eigenstates, the corresponding momenta are realized as
\be
p_\mu \sim -i\partial_\mu \ , \qquad  \bar p_{\bar\mu} \sim -i\partial_{\bar\mu}\ , \qquad \bar\psi^1_\mu\sim \frac{\partial}{\partial \psi^\mu_1}\ , \qquad \psi_{2\bar\mu}\sim\frac{\partial}{\partial \bar\psi^{2\bar\mu}} \ .
\ee
The states are therefore represented by functions $F(x,\bar x, \psi_1,\bar\psi^2)=\bra{x,\bar x, \psi_1,\bar\psi^2}F\rangle$, that, due to the Grassmannian nature of the $\psi$ variables, admit the finite Taylor expansion
\begin{equation}\label{F}
F(x,\bar x, \psi_1,\bar\psi^2) \ = \ \sum_{m,n=0}^d F_{\mu_1\ldots\mu_m,\bar\nu_1\ldots\bar\nu_n}(x,\bar x) \,\psi_1^{\mu_1}\ldots \psi_1^{\mu_m}\bar\psi^{2\bar\nu_1}\ldots \bar\psi^{2\bar\nu_n} \  ,
\end{equation}
where the coefficients $F_{\mu_1\ldots\mu_m,\bar\nu_1\ldots\bar\nu_n}(x,\bar x)$ are components of complex $(m,n)$-forms with $m$ antisymmetric holomorphic and $n$ antisymmetric anti-holomorphic indices. In the following, we shall sometimes use the shorthand notation $F_{\mu[m],\bar\nu[n]}(x,\bar x)$, where $m,n\in\mathbb{N}$ within square brackets denote the number of antisymmetrized indices of each kind.
Identifying $\psi_1^\m$ with the holomorphic differential $dx^\m$ and $\bar\psi^{2 \bar\m}$ with the anti-holomorphic one $d\bar x^{\bar\m}$, we can identify the various terms in the sum with $(m,n)$-forms
 \be
 F_{(m,n)}(x,\bar x) \ := \ F_{\mu[m],\bar\nu[n]}(x,\bar x)\,dx^{\m_1} \wedge \ldots \wedge dx^{\mu_m}\wedge d\bar x^{\bar\nu_1}\wedge \ldots \wedge d\bar x^{\bar\nu_n} \ .
 \ee
Wedge products will be understood in the following whenever differential forms are juxtaposed. On the above Hilbert-space states the quantized conserved charges are represented by differential operators. In particular, the supercharges $Q_1$ and $\bar Q^2$ act, up to an imaginary factor, as the Dolbeault operators $\del$ and $\bar\del$
\be\label{Q}
iQ_1 \ \sim \ \del \ = \ dx^\m\del_\m \ , \qquad i\bar Q^2 \ \sim \ \bar\del \ = \ d\bar x^{\bar\m}\del_{\bar\m} \ ,
\ee
on $(m,n)$-forms. Similarly,
\be\label{Qb}
i\bar Q^1 \ \sim \ -\del^\dagger \ = \ \delta^{\mu\bar\nu}\de_{\bar\nu}\frac{\del}{\del(dx^\m)} \ , \qquad i Q_2 \ \sim -\bar\del^\dagger \ = \ \delta^{\mu\bar\nu}\de_{\mu}\frac{\del}{\del(d\bar x^{\bar \nu})} \ ,
\ee
correspond, up to a sign, to the adjoint Dolbeault operators $\del^\dagger$ and $\bar\del^\dagger$ (see Appendix \ref{app:notations}) that act on $F_{(m,n)}$ by taking a divergence in the holomorphic or anti-holomorphic indices, respectively.
The hamiltonian is realized as the laplacian operator
\be
H \ \ \sim \ -\triangle \ := \ (\del\del^\dagger+\del^\dagger\del) \ = \ (\bar\del\bar\del^\dagger+\bar\del^\dagger\bar\del) \ = \ -\d^{\m\bar\n}\del_{\bar\n}\del_\m \ = \ - \frac{1}{2}\,\Box \ ,
\ee
where we denote with $\Box:=\del^M\del_M$ the laplacian operator in $\mathbb{R}^{2d}$
with its standard normalization.
Finally, the diagonal $R$-symmetry generators $J_1^1$ and $J_2^2$ essentially count the holomorphic and antiholomorphic degree of a $(m,n)$-form,
\bea
J_1^1 & = & N-\frac{d}{2} \ , \qquad  N \ \sim \ dx^\mu\frac{\de}{\de(dx^\mu)}\\
J_2^2 & = & -\bar N+\frac{d}{2} \ ,\qquad \bar N \sim \ d \bar x^{\bar\nu}\frac{\de}{\de(d\bar x^{\bar\nu})} \ ,
\eea
 respectively ($N$ and $\bar N$ correspond to the fermionic number operators in the holomorphic and anti-holomorphic sector); and the off-diagonal $R$-symmetry generators $J_1^2$ and $J_2^1$ correspond to inserting the metric and to taking a trace, respectively
 \be \label{gwedgeTr}
J_1^2 \ \sim \ \d\,\wedge \ = \ \d_{\mu\bar\nu}\,dx^\mu \,d\bar x^{\bar\nu} \ , \qquad J_2^1 \ \sim	\ -\Tr \ = \ -\d^{\mu\bar\nu}\,\frac{\de^2}{\de(dx^\mu)\de(d\bar x^{\bar\nu})} \ .
 \ee

We shall now proceed by gauging different subalgebras of the whole rigid supersymmetry algebra \eq{U2susyalg} generated by $C_I=(H, Q_i,\bar Q^i, J_i^j)$. The models we shall focus on are:
\begin{itemize}
\item $A$-model, obtained by gauging the hamiltonian $H$, the supercharges $Q_i$, $\bar Q^i$ and the $U(1)\times U(1)$ subgroup of the $R$-symmetry group $U(2)$, generated by $J^1_1$ and $J^2_2$. We denote the gauged generators by $C_I^A=(H, Q_i,\bar Q^i, J^1_1, J^2_2)$;
\item $\a$-model, obtained by gauging the whole $U(2)$-extended supersymmetry algebra \eq{U2susyalg}, \emph{i.e.} $C^\alpha_I=(H, Q_i,\bar Q^i, J_i^j)$;
\item $B$-model, obtained by leaving the supersymmetry rigid and gauging the hamiltonian $H$ and the $U(1)\times U(1)$ $R$-symmetry subgroup, $C_I^B=(H, J^1_1, J^2_2)$;
\item $\b$-model, obtained by gauging the hamiltonian and the whole $R$-symmetry algebra, \emph{i.e.} all generators \eq{charges} except for the supercharges, thus $C_I^\beta=(H, J_i^j)$.
\end{itemize}
The corresponding gauged worldline actions in phase space are obtained by coupling the correct subset of Noether charges $C_I$ to worldline gauge fields $G^I=(e, i\bar\chi^i, i\chi_i, a^i_j)$, where $e$ is the einbein, $\chi_i$, $\bar\chi^i$ are complex gravitinos, and $a^i_j$ is a worldline $U(2)$ gauge field.  By denoting the four models collectively as $\A=(A, \alpha, B, \beta)$, such actions read as
\begin{equation}\label{actions phase space}
S_{\A} = \int_0^1dt\Big[p_\mu\dot x^\mu+\bar p_{\bar\mu}\dot{\bar x}^{\bar\mu}+i\bar\psi^i_\mu\dot\psi_i^\mu-C_I^{\A}G^I\Big]+S^{CS}_{\A}\;,
\end{equation}
where $C_I^{\A}$ denotes the proper subset of charges that are gauged, while $S^{CS}_{\A}$ is the allowed Chern-Simons term for the $a$ gauge field.  When the whole $U(2)$ symmetry is gauged, only one Chern-Simons coupling $s$ is allowed
\begin{equation}
S^{CS}_{\alpha}=S^{CS}_{\beta} \ = \ s\int_0^1dt\,a^i_i\;,
\end{equation}
while, if only the subgroup $U(1)\times U(1)$ has to be gauged, one can add two independent couplings to the two $U(1)$ factors, $s_1$ and $s_2$,
\begin{equation}\label{CSA}
S^{CS}_{A}=S^{CS}_{B} \ = \ \int_0^1dt\Big[s_1\,a^1_1+s_2\,a^2_2\Big]\;.
\end{equation}

The classical equations of motion for the gauge fields $G^I$ constrain the Noether charges to vanish. Quantum-mechanically, this translates into the selection of the physical Hilbert space, which is obtained by requiring that the symmetry generators  annihilate physical states,
\be
\ket{F^{\A}}\in {\cal H}^{\A}_{phys} \qquad \Longleftrightarrow \qquad T_I^{\A}\ket{F^{\A}} \ = \ 0 \ , \quad \forall \ I \ ,
\ee
where we have defined (including the ordering issue discussed in eq. (\ref{ordering}))
\be
 T_I^{\A} \ := \  -\frac{\d S^{\A} }{\d G^I} \ = \  C_I^{\A} -\frac{\d S_{CS}^{\A} }{\d G^I}  \ .
\ee
Clearly, only the $J$-constraints are affected by the shift due to the Chern-Simons term,
 and take the form $T_J^{\A} = J_i^j-s\d_i^j$ for $\A=\a,\b$ and $T_J^{\A} = (J_1^1-s_1,J_2^2-s_2)$ for $\A=A,B$.

We now turn to examine the constraints in the different models and the resulting requirements they put on physical states.

\begin{itemize}
\item \emph{$\a$-model} \\[7pt]
We impose the diagonal $J$-constraints first. They take the form
\begin{equation}\label{Jiialpha}
\begin{split}
(J_1^1-s)\ket{F} & =  0 \ \ \longrightarrow \ \ (N-m)\sum_{k,l=0}^d F_{(k,l)} \ = \ 0  \ ,\\
(J_2^2-s)\ket{F} & =  0 \ \ \longrightarrow \ \  (-\bar N+d-m )\sum_{k,l=0}^d F_{(k,l)} \ = \ 0   \qquad m \ := \ s+\frac{d}2 \ \in \ \mathbb{N} \ .
\end{split}
\end{equation}
The restriction that $m$ be a natural number follows from the fact that $N$ and $\bar N$ count the number of holomorphic and anti-holomorphic indices, respectively. This requirement in turn fixes the possible quantized values of the Chern-Simons coupling $s$.  In summary, the $J_i^i$-constraints at fixed $i=1,2$ select, out of the various terms in \eq{F}, the $(p+1,d-p-1)$-form
\be \label{Falpha}
F_{(p+1,\,d-p-1)} \ = \ F_{\mu[p+1],\bar\nu[d-p-1]} \,dx^{\mu_1}\ldots dx^{\mu_{p+1}}d\bar x^{\bar\nu_1}\ldots d\bar x^{\bar\nu_{d-p-1}}  \ ,
\ee
as physical particle states in the $\a$-model, where we have set for convenience $m \equiv p+1$. Imposing now $J_2^1$ on \eq{Falpha} corresponds to declaring that
\be\label{Tr=0}
\Tr \,(F_{(p+1,\,d-1-p)}) \ = \ 0 \ .
\ee
While independent of $J_2^1$ at the level of the commutation relations \eq{U2susyalg}, the $J_1^2$ constraint is nonetheless automatically satisfied if \eq{Tr=0} holds, as one can show by letting $J_1^2$ act on \eq{Falpha}, using \eq{gwedgeTr} and the identity
\be
dx^{\m_1}\ldots dx^{\m_k}=\frac1{k!(d-k)!}\epsilon^{\m_1\ldots\m_k\n_{k+1}\ldots\n_d}\epsilon_{\r_1\ldots\r_k\n_{k+1}\ldots\n_{d}}dx^{\r_1}\ldots dx^{\r_k} \ ,
\ee
along with its anti-holomoprhic analogue. Let us stress that on a general K\"ahler manifold the (anti-)holomorphic volume form is not globally defined, unless the manifold is Calabi-Yau, \emph{i.e.} it has $SU(d)$ holonomy or, equivalently, admits a Ricci-flat metric. If this is not the case, the equivalence between $J_2^1$ and $J_1^2$ constraints only holds locally. We shall return to these issues in the forthcoming sections dealing with curved backgrounds.

As for the supercharges, it is immediate to see from the second relation in \eq{U2susyalg} that, once the $J$-constraints have been imposed, there are only two independent $(Q,\bar Q)$-constraints. We choose them to be $Q_1$ and $\bar Q^2$. From the realizations \eq{Q} and \eq{Qb} the conditions $Q_1\ket{F} = 0 = \bar Q^{2}\ket{F}$ correspond to (anti-)holomorphic integrability equations
\be
\del  F_{(p+1,\,d-1-p)} \ = \ 0 \ , \qquad \bar\del F_{(p+1,\,d-1-p)} \ = \ 0 \ ,
\ee
which are solved by
\be
F_{(p+1,\,d-1-p)} \ = \ \del\bar\del \a_{(p,\,d-2-p)} \ ,
\ee
where $\a_{(p,\,d-2-p)} $ is a potential represented by a complex $(p,d-2-p)$-form.  With such a position, the above equations become identities
(Bianchi identities). The spinning particle physical states therefore can be interpreted as curvatures for a complex gauge field $\a_{(p,\,d-2-p)} $.
Eq. \eq{Tr=0} now becomes a  second-order wave equation
\be\label{eomalpha}
\Tr \,(\del\bar\del \a_{(p,\,d-2-p)} ) \ = \ \left(\triangle+\del\del^\dagger+\bar\del\bar\del^\dagger-\del\bar\del\Tr\right)\a_{(p,\,d-2-p)}  \ = \ 0 \ ,
\ee
which just corresponds to a complexified version of Fronsdal's equations \cite{Fronsdal:1978rb, Bastianelli:2009vj}. Indeed, in components it reads
\begin{equation}
\begin{split}
\del_\r\bar\del^\r\a_{\m[p],\,\bar\n[d-2-p]} &-p\,\del_{[\m_1|}\bar\del^\r\a_{\r|\m_2\ldots\m_p],\,\bar\n[d-2-p]} -(d-2-p)\,\del_{[\bar\n_1|}\del^{\bar\r}\a_{\m[p],\,\bar\r|\bar\n_2\ldots\bar\n_{d-2-p}]}\\[2mm]
&+p(d-2-p)\,\del_{[\m_1}\del_{[\bar\n_1|}\a^{\bar\s}{}_{\m_2\ldots\m_p],\,\bar\s|\bar\n_2\ldots\bar\n_{d-2-p}]} \ = \ 0 \ ,
\end{split}
\end{equation}
where we denoted weighted anti-symmetrization with square brackets.
Clearly, the curvatures and the equations of motion \eq{eomalpha} are invariant under the gauge transformations
\be
\d\a_{(p,\,d-2-p)} \ = \ \del\L^1_{(p-1,\,d-2-p)}+\bar\del\L^2_{(p,\,d-3-p)} \ .
\ee
As we mentioned in the introduction, the $\alpha$-form system discussed above is physically equivalent to the bi-forms described in \cite{Bastianelli:2009vj} for $N=2$, since they come from the very same quantum mechanical sigma model. In the former work it has been used a Hilbert space basis of $(\psi_1,\psi_2)$ eigenstates, instead of the present basis of $(\psi_1,\bar\psi^2)$ eigenstates. This leads to different geometric realizations of the same quantum mechanical operators, the curvatures in the two realizations being related by a holomorphic Hodge duality as
$$
F^{\text{ new}}_{\mu[m], \bar\nu_1...\bar\nu_{d-m}}\propto F^{\text{ old}}_{\mu[m],\nu_1...\nu_m}\,\epsilon^{\nu_1...\nu_m}{}_{\bar\nu_1...\bar\nu_{d-m}}\;,
$$
and gauge fields are introduced by solving different integrability equations that give rise to different Bianchi identities.
Hence, gauge potentials in the two realizations are not related in a local way, and the corresponding field theories are not manifestly equivalent at the level of equations of motion.

\item \emph{$A$-model}\\[7pt]
The difference with the $\a$-model consists in the absence of the non-diagonal $J$-constraints. As a consequence, as explained in \eq{CSA}, here we have two independent Chern-Simons couplings, one for each $U(1)$. In other words, gauging $(J_1^1,J_2^2) $ enforces the requirements
\begin{equation}\label{JiiA}
\begin{split}
(J_1^1-s_1)\ket{F} & =  0 \ \ \longrightarrow \ \ (N-m)\sum_{k,l=0}^d F_{(k,l)} \ = \ 0\ , \quad m \ := \ s_1+\frac{d}2 \ \in \ \mathbb{N}  \ ,\\
(J_2^2-s_2)\ket{F} & =  0 \ \ \longrightarrow \ \  (-\bar N+n )\sum_{k,l=0}^d F_{(k,l)} \ = \ 0   \quad n \ := \ -s_2+\frac{d}2 \ \in \ \mathbb{N} \ ,
\end{split}
\end{equation}
that select physical states represented by $(p+1,q+1)$-forms
\be \label{FA}
F_{(p+1,\,q+1)} \ = \ F_{\mu[p+1],\bar\nu[q+1]} \,dx^{\mu_1}\ldots dx^{\mu_{p+1}}d\bar x^{\bar\nu_1}\ldots d\bar x^{\bar\nu_{q+1}}  \ ,
\ee
where we have set $m\equiv p+1$, $n\equiv q+1$. Now that $J_2^1$ and $J_1^2$ are no longer imposed, all supersymmetry generators give rise to independent constraints. As before, $Q_1$ and $\bar Q^2$ enforce the integrability equations
\be
\del  F_{(p+1,\,q+1)} \ = \ 0 \ , \qquad \bar\del F_{(p+1,\,q+1)} \ = \ 0 \ ,
\ee
that are solved in terms of a $(p,q)$-form potential $A_{(p,\,q)}$,
\be\label{curvpq}
F_{(p+1,\,q+1)} \ = \ \del\bar\del A_{(p,\,q)} \ .
\ee
The remaining constraints $\bar Q^1\ket{F}  =0=Q_2\ket{F}$ translate into
\bea
\del^\dagger (\del\bar\del A_{(p,\,q)}) & = & \bar\del(-\triangle A_{(p,\,q)}-\del\del^\dagger A_{(p,\,q)}) \ = \ 0 \ ,\\
\bar\del^\dagger (\del\bar\del A_{(p,\,q)}) & = & \del(\triangle A_{(p,\,q)}+\bar\del\bar\del^\dagger A_{(p,\,q)}) \ = \ 0 \ ,
\eea
from which, taking into account that $\del^2=0=\bar\del^2$, it follows that $(\triangle+\del\del^\dagger+\bar\del\bar\del^\dagger)A_{(p,\,q)}\in \textrm{Ker}(\del,\bar\del)$. In other words, the field equations
\be\label{eomA}
\left(\triangle+\del\del^\dagger+\bar\del\bar\del^\dagger\right)A_{(p,\,q)} \ = \ \del\bar\del\r_{(p-1,\,q-1)} \ ,
\ee
where $\r_{(p-1,\,q-1)}$ is a compensator field, are satisfied by the $(p,q)$-form potential $A_{(p,\,q)}$ characterizing the physical spinning particle states in the $A$-model via \eq{curvpq}. Eq. \eq{eomA} is a complex version of the Francia-Sagnotti equations adapted to a $(p,q)$-form potential
\cite{Francia:2002aa,Francia:2002pt}.
A geometrical derivation of similar Maxwell-like
equations with compensators leading to expressions
close in form to (\ref{eomA}) was given in \cite{Francia:2010qp}.
The curvature \eq{curvpq} is left invariant by the gauge transformations
\be
\d A_{(p,\,q)} \ = \ \del\L^1_{(p-1,\,q)}+\bar\del\L^2_{(p,\,q-1)} \ .
\ee
This transformation preserves the field equations \eq{eomA} provided that it is accompanied by the gauge variation of the compensators
\be\label{delta rho}
\d\r_{(p-1,\,q-1)}  \ = \ \bar\del^\dagger\L^1_{(p-1,\,q)}-\del^\dagger\L^2_{(p,\,q-1)} \ ,
\ee
as it can be proved by direct substitution.
It is possible, by means of \eqref{delta rho}, to gauge fix the compensators to zero, recovering the field equations in the homogeneous form $(\triangle+\del\del^\dagger+\bar\del\bar\del^\dagger)A_{(p,\,q)}=0$. The residual gauge symmetry parameters are then forced to satisfy $\bar\del^\dagger\L^1_{(p-1,\,q)}-\del^\dagger\L^2_{(p,\,q-1)}=0$, analogously to the case studied in \cite{Bastianelli:2009vj}. Moreover, by taking divergencies of the gauge fixed field equations, one finds that the $(p,q)$-form has to be doubly divergenceless: $\de^\dagger\bar\de^\dagger A_{(p,q)}=0$.

\item \emph{$\b$-model}\\[7pt]
In this case the hamiltonian and the whole $U(2)$ $R$-symmetry algebra are gauged, while the supersymmetries are not. This means that the $J$-constraints \eqref{Jiialpha}-\eq{Tr=0} are all imposed, but there are no integrability equations to be solved in terms of potentials, nor there are Fronsdal-like equations for the latter. This implies that the $(m,n)$-forms $F_{(m,n)}$ that enter the expansion \eq{F} have no potential and no associated gauge symmetry. For this reason, we shall here set $m \equiv p$ for a direct comparison with the previous models and write the $(m,n)$-forms selected by \eqref{Jiialpha} as $\b_{(p,\,d-p)}$.
Therefore, the physical states in this model are represented by traceless $(p,d-p)$-forms $\b_{(p,\,d-p)}$ satisfying a Klein-Gordon equation
\be\label{eombeta}
\triangle \beta_{(p,\,d-p)} \ = \ 0\ ,\qquad \Tr(\beta_{(p,\,d-p)}) \ = \ 0\ .
\ee

\item \emph{$B$-model}\\[7pt]
Again the supersymmetries remain rigid and the only $R$-symmetry constraints that are imposed are \eqref{JiiA}. As a result, in the $B$-model the physical states are represented by $(p,q)$-forms $B_{(p,q)}$ satisfying the Klein-Gordon equation
\be\label{eomB}
\triangle B_{(p,\,q)} \ = \ 0
\ee
with no associated gauge symmetries.

\end{itemize}
Though derived in flat space, all these equations, suitably covariantized, hold on a generic K\"ahler manifold, as the analysis in the following section will show.

\section{$(p,q)$-forms on K\"ahler manifolds}\label{sec curved space}

We shall now analyze the coupling of the various $(p,q)$-forms considered above to an arbitrary background K\"ahler metric $g_{\m\bar\n}(x,\bar x)=g_{\bar\n\m}(x,\bar x)$. The Grassmann variables $(\psi_i^\m(t),\bar\psi^i_\m(t))$ now transform as contravariant or covariant vectors under holomorphic change of coordinates, and we stress that in curved space we choose to define the $\bar\psi$'s with an holomorphic lower vector index, in order to avoid position dependent anti-commutation relations\footnote{These variables correspond to Darboux coordinates, that make the graded sympletic form assume a canonical expression
with constant components. By canonical quantization these coordinates acquire then position independent anti-commutation relations.}.
Accordingly, appropriate covariantizations of the supersymmetry charges need to be constructed, such that their anticommutation relations produce the correct supersymmetry algebra and related hamiltonian.
Recalling that on K\"ahler manifolds the only non-vanishing components of the Christoffel connection are the purely holomorphic $\C^\l_{\m\n}$ and purely anti-holomorphic ones $\C^{\bar\l}_{\bar\m\bar\n}$, one can substitute the momenta $(p_\m,\bar p _{\bar\m})$ with the covariant momenta $(\pi_\m,\bar \pi _{\bar\m})$ defined as
\be\label{covpi}
\pi_\m \ = \ p_\m+i\,\C^\l_{\m\n}\,\psi^\n_i\bar\psi^i_\l  \ , \qquad \bar\pi_{\bar\m} \ = \ \bar p _{\bar\m} \ ,
\ee
whose Poisson bracket is proportional to the Riemann curvature tensor (see Appendix \ref{app:notations} for our curvature conventions)
\be
\left\{\pi_\m,\bar \pi _{\bar\n}\right\}_{PB} \ = \ iR_{\m\bar\n}{}^\l{}_\n\,\psi^\n_i\bar\psi^i_\l \ .
\ee
Therefore, the classical covariantized supercharges read
\bea\label{Q cov}
Q_i & = & \psi^\m_i\,\pi_\m , \qquad \bar Q^i \ = \ \bar\psi^i_\m \,g^{\m\bar\n}\,\bar p_{\bar\n}\ ,
\eea
and their Poisson bracket leads to the classical hamiltonian
\be\label{Hcl}
\begin{split}
H_{cl} & = \ g^{\m\bar\n}\,\bar\pi_{\bar\n}\pi_\m -\frac{1}{2}R_\mu{}^\nu{}_\lambda{}^\sigma\,\psi^\mu_i\bar\psi_\nu^i\psi^\lambda_j\bar\psi_\sigma^j\\
& = \ g^{\m\bar\n}\,\bar p_{\bar\n}( p_\m+i\,\C^\l_{\m\n}\,\psi^\n_i\bar\psi^i_\l) -\frac{1}{2}R_\mu{}^\nu{}_\lambda{}^\sigma\,\psi^\mu_i\bar\psi_\nu^i\psi^\lambda_j\bar\psi_\sigma^j\ .
\end{split}
\ee
The classical phase space actions for the various models can now be written in terms of the covariant momenta as
\bea\label{covactions phase space}
S_{\A} & = & \int_0^1dt\Big[p_\mu\dot x^\mu+\bar p_{\bar\mu}\dot{\bar x}^{\bar\mu}+i\bar\psi^i_\mu\dot\psi_i^\mu-{\cal C}_I^{\A}{G}^I\Big]+S^{CS}_{\A}
\eea
where ${\cal C}_I^{\A}$ differs from $C_I^{\A}$ of the previous section for the substitution of the flat space constraints with the covariantized ones \eqref{Q cov}, \eqref{Hcl}.

At the quantum level an ordering prescription is again needed for the $U(d)$ generators $\psi^\m_i\bar\psi^i_\l$ appearing in the covariant momenta \eq{covpi}. We therefore set
\be\label{covpi quantum}
\pi_\m \ = \ p_\m+i\,\C^\l_{\m\n}\,M^\n_\l  \ , \qquad \bar\pi_{\bar\m} \ = \ \bar p _{\bar\m} \ ,
\ee
where the $U(d)$ ``Lorentz'' generators
\be\label{M}
M^\m_\n  \ := \ \frac{1}{2}\left[\psi^\m_i,\bar\psi^i_\n\right] \ = \ \psi^\m_i\bar\psi^i_\n-\d^\m_\n
\ee
(see also Eq. \eqref{Ud}) are here defined with the graded-symmetric ordering prescription, and their commutator reads
\be
\left[\pi_\m,\bar \pi _{\bar\n}\right] \ = \ -R_{\m\bar\n}{}^\l{}_\n\,\,M^\n_\l \ .
\ee
The covariant momenta are hermitian conjugate to each other with respect to the inner product
\be\label{innerg}
\bra{\chi}\phi\rangle \ = \ \int d^dx\, d^d\bar x \,g(x,\bar x) \,d^d\psi \,d^d\bar\psi \ e^{\bar\psi_\m^i \psi_i^\m}\, \overline{\chi(x,\bar x,\psi_1,\bar\psi^2)}\phi(x,\bar x,\psi_1,\bar\psi^2)
\ee
where $g=\det g_{\m\bar\n}$. Note that with this inner product $(\psi_i^\m)^\dagger=\bar\psi^i_\n g^{\bar\m\n}$, and the hermiticity property of the momentum reads: $p_\m^\dagger=\bar p_{\bar\m}+ig_{\l\bar\l}g^{\n\bar\n}\,\C^{\bar\l}_{\bar\m\bar\n}\,M^\l_\n$.  The supercharges then take the form
\bea
Q_i & = & \psi^\m_i\,g^{1/2}\,\pi_\m\,g^{-1/2} \ , \\[1mm]
\bar Q^i & = & \bar\psi^i_\m \,g^{\m\bar\n}\,g^{1/2}\,\bar\pi_{\bar\n}\,g^{-1/2} \ ,
\eea
where the wrappings $g^{1/2}\ldots g^{-1/2} $ ensure their correct hermiticity properties \cite{Bastianelli:2009vj, Bastianelli:2011pe} since, with respect to the inner product \eq{innerg}, one has $Q^\dagger_i \ = \ \bar Q^i$. The covariantized supercharges satisfy the anti-commutation relation
\bea
\{Q_i,\bar Q^j\} & = & g^{1/2}\{\psi^\m_i\pi_\m,\bar\psi^i_\m \,g^{\m\bar\n}\,\bar\pi_{\bar\n}\}g^{-1/2}\nn \\
& = & \d_i^j g^{1/2}g^{\m\bar\n}\bar\pi_{\bar\n}\pi_\m g^{-1/2} -\psi_i^{\m}\bar\psi^{j\bar\n}R_{\m\bar\n\s\bar\l}M^{\s\bar \l} \nn\\
& = & \d_i^j \left(g^{1/2}g^{\m\bar\n}\bar\pi_{\bar\n}\pi_\m g^{-1/2} -\frac{1}{2}R_{\m\bar\n\s\bar\l}M^{\m\bar\n}M^{\s\bar \l}+\frac{1}{2}R_{\m\bar\n}M^{\m\bar\n} \right)\nn \\
&=& \d_i^j H\;,
\eea
where the fact that $N=2$ is crucial for the next-to-last equality, and where we have defined
\bea
H & := &  H_0-\frac{1}{2}R_{\m\bar\n\s\bar\l}M^{\m\bar\n}M^{\s\bar \l}  \\
H_0 & := & \frac{1}{2} g^{1/2}g^{\m\bar\n}(\bar\pi_{\bar\n}\pi_\m
+\pi_\m\bar\pi_{\bar\n})g^{-1/2}\ .
\eea
The other commutators \eq{U2susyalg} remain the same as in the flat case. One therefore achieves closure of the constraint algebra with the complete hamiltonian $H$, containing the symmetric, minimally-covariantized piece $H_0$ and a non-minimal contribution proportional to the Riemann curvature. A point worth stressing is that the closure of the $(Q_i,\bar Q^j)$ algebra on $H$ forbids an extra coupling to the $U(1)$ part of the K\"ahler connection, differently from the case of $(p,0)$-forms treated in \cite{Bastianelli:2011pe}, for which an arbitrary $U(1)$ charge is allowed.
With this restriction one sees that performing (anti-)holomorphic Hodge dualities with the chiral epsilon tensor $e\,\epsilon_{\mu_1...\mu_d}$ would lead us outside the  class of
models described by our spinning particle, since unwanted couplings to the $U(1)$ part of the holonomy would be introduced\footnote{We denote with $e$ the determinant of the vielbein $e_\mu^a$. See Appendix B in \cite{Bastianelli:2011pe} for details.}.

Having presented the covariant model on a K\"ahler manifold, it is now useful to spend some words on the geometric realization of the operators and the Hilbert space states. Since we are expanding the states in powers of $\psi_1^\mu$ and $\bar\psi^2_\mu$, our tensors contain only holomorphic indices, and take the form
\begin{equation}\label{Fcurved}
F(x,\bar x, \psi_1,\bar\psi^2) \ = \ \sum_{m,n=0}^d F_{\mu_1\ldots\mu_m}^{\nu_1\ldots\nu_n}(x,\bar x) \,\psi_1^{\mu_1}\ldots \psi_1^{\mu_m}\bar\psi^{2}_{\nu_1}\ldots \bar\psi^{2}_{\nu_n} \  .
\end{equation}
These tensors are of course equivalent to the $(m,n)$-forms by raising  the antiholomorphic indices with the K\"ahler metric
\begin{equation}\label{form vs tensors}
F_{\mu_1\ldots\mu_m}^{\nu_1\ldots\nu_n}=g^{\nu_1\bar\nu_1}\ldots g^{\nu_n\bar\nu_n}\,F_{\mu_1\ldots\mu_m,\bar\nu_1\ldots\bar\nu_n}\;.
\end{equation}
The covariant momenta \eqref{covpi quantum} acts on the tensors in \eqref{Fcurved} as covariant derivatives:
\begin{equation}\label{covariant momenta}
g^{1/2}\,\pi_\mu\, g^{-1/2}\sim-i\nabla_\mu\;,\quad
g^{1/2}\,\bar\pi_{\bar\mu}\, g^{-1/2}\sim-i\de_{\bar\mu}=-i\nabla_{\bar\mu}\;,
\end{equation}
and only when the operators are expressed as covariant geometric objects one is free to pass through the $g$ factors in \eqref{form vs tensors}, and let them act on the true $(m,n)$-form fields. Hence, once the covariant action of geometric operators is understood, it is possible to identify the $\bar\psi^2$'s as anti-holomorphic form basis via $g^{\mu\bar\nu}\bar\psi^2_\mu\sim d\bar x^{\bar\nu}$, and the conjugate momenta as formal derivatives thereof: $\psi_2^\mu\sim g^{\mu\bar\nu}\,\frac{\de}{\de(d\bar x^{\bar\nu})}$. To make more explicit how this is realized, we write the action of the covariant momentum $\bar\pi_{\bar\mu}=\bar p_{\bar\mu}$ on a state in \eqref{Fcurved}
\begin{equation}
\begin{split}
&ig^{1/2}\,\bar\pi_{\bar\mu}\,g^{-1/2}\,F_{\mu_1...\mu_m}^{\nu_1...\nu_n} \,\psi_1^{\mu_1}... \psi_1^{\mu_m}\bar\psi^{2}_{\nu_1}... \bar\psi^{2}_{\nu_n}\\
&=\Big[\de_{\bar\mu}\,F_{\mu_1...\mu_m}^{\nu_1...\nu_n}\Big] \,\psi_1^{\mu_1}... \psi_1^{\mu_m}\bar\psi^{2}_{\nu_1}... \bar\psi^{2}_{\nu_n}\\
&=\Big[\nabla_{\bar\mu}F_{\mu_1...\mu_m}^{\nu_1...\nu_n}\Big] \,\psi_1^{\mu_1}... \psi_1^{\mu_m}\bar\psi^{2}_{\nu_1}...\bar\psi^{2}_{\nu_n}\\
&=g^{\nu_1\bar\nu_1}... g^{\nu_n\bar\nu_n}\,\Big[\nabla_{\bar\mu}F_{\mu_1...\mu_m,\bar\nu_1...\bar\nu_n}\Big]\,\psi_1^{\mu_1}... \psi_1^{\mu_m}\bar\psi^{2}_{\nu_1}... \bar\psi^{2}_{\nu_n}\\
&=\nabla_{\bar\mu}F_{\mu_1...\mu_m,\bar\nu_1...\bar\nu_n}\,dx^{\mu_1}... dx^{\mu_m}d\bar x^{\bar\nu_1}... d\bar x^{\bar\nu_n}\;,
\end{split}
\end{equation}
where we used \eqref{form vs tensors} and identified $g^{\mu\bar\nu}\bar\psi^2_\mu\sim d\bar x^{\bar\nu}$. Let us notice that in the last two lines the covariant derivative $\nabla_{\bar\mu}$ contains the needed connections $\Gamma^{\bar\l}_{\bar\n\bar\s}$ acting on the anti-holomorphic indices. We see then that the identification \eqref{covariant momenta} is valid, keeping in mind the discussion above, also at the level of $(m,n)$-forms, and only when acting on $(m,n)$-forms the supercharges correctly reproduce the Dolbeault operators and their adjoints
\begin{equation}\label{QQbar}
\begin{split}
& iQ_1 \ \sim \ \del \ = \ dx^\m\del_\m \ , \qquad i\bar Q^2 \ \sim \ \bar\del \ = \ d\bar x^{\bar\m}\bar\del_{\bar\m} \ ,\\
& i\bar Q^1 \ \sim \ -\del^\dagger \ = \ \bar\nabla^\mu\frac{\del}{\del(dx^\m)} \ , \qquad i Q_2 \ \sim -\bar\del^\dagger \ = \ \nabla^{\bar\mu}\frac{\del}{\del(d\bar x^{\bar \mu})} \ ,
\end{split}
\end{equation}
with $\bar\nabla^\mu=g^{\mu\bar\nu}\nabla_{\bar\nu}$ and $\nabla^{\bar\mu}=g^{\nu\bar\mu}\nabla_{\nu}$.
The quantum $U(d)$ generators \eq{M} act as
\begin{equation}
M^{\mu\bar\nu} \ \sim \ g^{\nu\bar\nu}\,dx^\mu\frac{\de}{\de(dx^\nu)}-g^{\mu\bar\mu}\,d\bar x^{\bar\nu}\frac{\de}{\de(d\bar x^{\bar\mu})}\ ,
\end{equation}
while the complete hamiltonian corresponds to the Laplace-Beltrami operator
\bea\label{Htot}
H & \sim & -\triangle \ = \ \del\del^\dagger+\del^\dagger\del \ = \ \bar\del\bar\del^\dagger+\bar\del^\dagger\bar\del \\
& = &- \frac{\nabla^2}{2}-\frac12\,R_{\mu\bar\nu\lambda\bar\sigma}\,M^{\mu\bar\nu}M^{\lambda\bar\sigma} \ ,
\eea
where $\nabla^2$ is the $2d$-dimensional curved space laplacian
\begin{equation}
\nabla^2=G^{MN}\nabla_M\nabla_N=g^{\mu\bar\nu}\left(\nabla_\mu\nabla_{\bar\nu}+\nabla_{\bar\nu}\nabla_\mu\right)\ .
\end{equation}
Finally, the action of the off-diagonal $R$-symmetry generators \eq{gwedgeTr} now corresponds to the insertion of the K\"ahler form (up to an imaginary factor) and to a trace with respect to the K\"ahler metric, respectively,
 \be \label{KgwedgeTr}
J_1^2 \ \sim \ g\,\wedge \ = \ g_{\mu\bar\nu}\,dx^\mu \,d\bar x^{\bar\nu} \ , \qquad J_2^1 \ \sim 	\ -\Tr \ = \ -g^{\mu\bar\nu}\,\frac{\de^2}{\de(dx^\mu)\de(d\bar x^{\bar\nu})} \ .
 \ee
We stress here that, since the holomorphic volume form $e\,\epsilon_{\mu_1...\mu_d}$ is not globally defined, unless the manifold is Calabi-Yau, the equivalence between the $J_1^2$ and $J_2^1$ constraints is affected by topological mismatches  that  appear when treating the $\alpha$ forms,
as already anticipated in Sec. \ref{sec flat space}.

Since the extended supersymmetry algebra has the same form as in flat space, the field equations of the four models keep the same form \eq{eomalpha}, \eq{eomA}, \eq{eombeta}, and \eq{eomB} in terms of the operators $\de,\bar\de,\de^\dagger,\bar\de^\dagger,\triangle, \Tr$. The difference consists in replacing $\d_{\m\bar\n}$ with $g_{\m\bar\n}$ and their inverses, in the covariant $(\de^\dagger,\bar\de^\dagger)$ operators given by \eqref{QQbar}, and in the full hamiltonian \eq{Htot}, realized as the full Laplace-Beltrami operator that also contains curvature corrections: $\triangle=\frac{\nabla^2}{2}+\frac12\,R_{\mu\bar\nu\lambda\bar\sigma}\,M^{\mu\bar\nu}M^{\lambda\bar\sigma}$.

\section{Effective action of differential forms}\label{sec effective actions}

In the present section we wish to study the functional quantization of our spinning particle models coupled to a curved K\"ahler background. This will provide the heat kernel expansion of the effective actions for the different field theories of $(p,q)$-forms, as well as exact relations between Hodge dual descriptions of the same quantum theory. The proofs of exact dualities and the related topological issues will be carefully studied in the next section.

The effective action for such differential forms on K\"ahler spaces is recovered in the worldline formalism by quantizing the corresponding spinning particles on a
circle (a one-dimensional torus). The worldline actions in phase space are obtained by coupling the correct subset of Noether charges $\mathcal{C}^\A_I$, listed in the previous section for the different models, to worldline gauge fields, and are given by \eqref{covactions phase space}.
By eliminating momenta in \eqref{covactions phase space} and performing a Wick rotation, we recover the euclidean actions in configuration space for the four theories, that explicitly read
\begin{equation}\label{actions conf space}
\begin{split}
S_A &= \int_0^1 d\tau\,\Big[e^{-1}g_{\mu\bar\nu}\big(\dot x^\mu-\bar\chi^i\psi_i^\mu\big)\big(\dot{\bar x}^{\bar\nu}
-\chi_i\bar\psi^{i\bar\nu}\big)+\bar\psi^1_\mu\big[D_{\tau}+ia^1_1\big]\psi_1^\mu+is_1\,a_1^1\\
&+\bar\psi^2_\mu\big[D_{\tau}+ia^2_2\big]\psi_2^\mu+is_2\,a^2_2-\frac{e}{2}R_\mu{}^\nu{}_\lambda{}^\sigma\,\psi^\mu_i\bar\psi_\nu^i\psi^\lambda_j\bar\psi_\sigma^j\Big]\\
S_\alpha &= \int_0^1 d\tau\,\Big[e^{-1}g_{\mu\bar\nu}\big(\dot x^\mu-\bar\chi^i\psi_i^\mu\big)\big(\dot{\bar x}^{\bar\nu}
-\chi_i\bar\psi^{i\bar\nu}\big)+\bar\psi^i_\mu\big[\delta^j_iD_{\tau}+ia^j_i\big]\psi_j^\mu+is\,a_i^i\\
&-\frac{e}{2}R_\mu{}^\nu{}_\lambda{}^\sigma\,\psi^\mu_i\bar\psi_\nu^i\psi^\lambda_j\bar\psi_\sigma^j\Big]\\
S_B &= \int_0^1 d\tau\,\Big[e^{-1}g_{\mu\bar\nu}\dot x^\mu\dot{\bar x}^{\bar\nu}+\bar\psi^1_\mu\big[D_{\tau}+ia^1_1\big]\psi_1^\mu+is_1\,a_1^1
+\bar\psi^2_\mu\big[D_{\tau}+ia^2_2\big]\psi_2^\mu+is_2\,a^2_2\\
&-\frac{e}{2}R_\mu{}^\nu{}_\lambda{}^\sigma\,\psi^\mu_i\bar\psi_\nu^i\psi^\lambda_j\bar\psi_\sigma^j\Big]\\
S_\beta &= \int_0^1 d\tau\,\Big[e^{-1}g_{\mu\bar\nu}\dot x^\mu\dot{\bar x}^{\bar\nu}+\bar\psi^i_\mu\big[\delta^j_iD_{\tau}+ia^j_i\big]\psi_j^\mu+is\,a_i^i-\frac{e}{2}R_\mu{}^\nu{}_\lambda{}^\sigma\,\psi^\mu_i\bar\psi_\nu^i\psi^\lambda_j\bar\psi_\sigma^j\Big]\;,
\end{split}
\end{equation}
where the covariant time derivative is given by $D_\tau\psi_i^\mu=\dot\psi^\mu_i+\dot x^\nu\Gamma^\mu_{\nu\lambda}\,\psi^\lambda_i$. Note that along with the Wick rotation $t\to-i\tau$ we have also rotated the gauge fields $a^i_j\to -ia^i_j$ to keep the gauge group compact.

The quantization of the spinning particle on a circle parameterized by the proper time $\tau\in[0,1]$, gives the effective action for the differential forms we are interested in, coupled to the background metric $g_{\mu\bar\nu}(x,\bar x)$
\begin{equation}\label{partition function defined}
Z_{\A}[g]\propto\int\left[\frac{\Dc X\Dc G}{\text{Vol(Gauge)}}\right]_{\A}\, {\rm e}^{-S_{\A}[X,G_\A]}\;.
\end{equation}
We denote the dynamical variables as $X=(x^\mu, \bar x^{\bar\mu}, \psi_i^\mu, \bar\psi_\mu^i)$, while $G_\A$ is the subset of $(e, \bar\chi^i, \chi_i, a^i_j)$ needed for the model $\A$. The subscript $\A$ in the functional measure stands for the dependence of the measure itself on the choice of the model, \emph{i.e.}, on the choice of the worldline gauge group.

By means of the Faddeev-Popov procedure, we gauge fix the ``supergravity multiplet'' to the constant values
$$
\tilde G=\left(\beta,0,0,\left(
                       \begin{array}{cc}
                         \phi & 0 \\
                         0 & \theta \\
                       \end{array}
                     \right)
\right)\;,
$$
and we are left with modular integrations over $\beta$, $\phi$ and $\theta$, with the one-loop measure that was carefully studied in \cite{Bastianelli:2009vj, Bastianelli:2007pv}
\begin{equation}\label{partition function}
Z_\A[g]\propto\int_0^\infty\frac{d\beta}{\beta}\int_0^{2\pi}\frac{d\phi}{2\pi}\int_0^{2\pi}\frac{d\theta}{2\pi}\,\tilde\mu_\A(\phi,\theta)\int_{\text{P}}\Dc x\Dc\bar x\int_{\text{A}}
D\bar\psi D\psi\, {\rm e}^{-S_\A[X,\tilde G_\A]}
\end{equation}
where $S[X,\tilde G_\A]$ stands for the gauge fixed action, \emph{i.e.} \eqref{actions conf space}
evaluated at $G=\tilde G$. The subscript P and A denote periodic and antiperiodic boundary conditions, respectively.
The integral over $\beta$ is the usual proper time integral with the well known one-loop measure.
The term $\tilde\mu_\A(\phi, \theta)$ contains the Faddeev-Popov factors, depending on the chosen model: when the supersymmetry is gauged ($A$ and $\alpha$ models), the bosonic superghosts associated to $\chi_i$ and $\bar\chi^i$ produce a factor $\Big(2\cos\frac\phi2\,2\cos\frac\theta2\Big)^{-2}$, and when the whole $U(2)$ $R$-symmetry is gauged ($\alpha$ and $\beta$ models), the ghosts of the non-abelian group produce the
additional term $\frac12\Big(2\sin\frac{\phi-\theta}{2}\Big)^2$, see \emph{e.g.} \cite{Bastianelli:2009vj,Bastianelli:2007pv}.

We denote with $\Dc x$ the general coordinate invariant measure, \emph{i.e.}
$$
\Dc x \Dc \bar x \sim\prod_{\tau=0}^1d^dx(\tau)\, d^d\bar x(\tau)\, g(x(\tau), \bar x(\tau))\;,
$$
with $g=\det g_{\mu\bar\nu}$, while $D\psi \sim\prod_{\tau=0}^1d^d\psi(\tau)$
is the translational invariant flat measure.\footnote{Note that, since $\psi$'s are spacetime vectors, while $\bar\psi$'s are covectors,
 the covariant measure coincides with the flat one: $\Dc\bar\psi\Dc\psi=D\bar\psi D\psi$.} Formula \eqref{partition function} gives the worldline representation of the effective action for the differential form $\A$. At this stage, all the dependence on the model chosen is contained in the modular measure $\tilde\mu_\A(\phi, \theta)$ and in the Chern-Simons part of the action. To exploit it further, we need some more manipulations. In order to extract the integral over spacetime, we choose an arbitrary $x_0$ as a base-point for our loops.
The periodic path integral then factorizes as $\int_{\text{P}}\Dc x\Dc\bar x=\int d^dx_0d^d\bar x_0g(x_0)\int_{x(0)=x(1)=x_0}\!\!\!\!\Dc x\Dc\bar x$,
and we let the coordinate fields fluctuate around the fixed point $x_0$ as: $x^\mu(\tau)=x_0^\mu+q^\mu(\tau)$, with $q^\mu(0)=q^\mu(1)=0$.
The $x$ path integral becomes $\int_{\text{D}}\Dc q\Dc\bar q$, where D stands for Dirichlet boundary conditions. The remaining obstacle in performing a perturbative expansion is the field dependent measure $\Dc q\Dc\bar q$. Following \cite{Bastianelli:1991be, Bastianelli:1992ct} we
exponentiate the metric determinant with a path integral over fermionic complex ghosts $b^\mu$ and $\bar c^{\bar\nu}$:
$\Dc q\Dc\bar q=DqD\bar q\int DbD\bar c\; {\rm e}^{-S_{\text{gh}}}$. The full gauge fixed action, containing the ghost term: $S_\A^{\text{gf}}\equiv S_\A[X,\tilde G_\A]+S_{\text{gh}}$ takes the following form\footnote{We rescaled fermions by $\psi\to \frac{1}{\sqrt\beta}\psi$ to have common normalizations of the two-point functions.}
\begin{equation}
\begin{split}
S_\A^{\text{gf}} &= \frac1\beta\int_0^1d\tau\,\Big[g_{\mu\bar\nu}\big(\dot q^\mu\dot{\bar q}^{\bar\nu}
+b^\mu\bar c^{\bar\nu}\big)+\bar\psi^1_\mu(D_\tau+i\phi)\psi^\mu_1+\bar\psi^2_\mu(D_\tau+i\theta)\psi^\mu_2\\
&-\frac{1}{2}R_\mu{}^\nu{}_\lambda{}^\sigma\,\psi^\mu_i\bar\psi_\nu^i\psi^\lambda_j\bar\psi_\sigma^j
\Big]+\tilde S_\A^{CS} \;,
\end{split}
\end{equation}
where $\tilde S_\A^{CS}$ is the gauge fixed euclidean Chern-Simons term, that we choose to plug in the measure $\tilde\mu_\A$.

In order to perform perturbative calculations we expand all background fields around the fixed point $x_0$.
The action written above splits into a quadratic part $S_2$ giving propagators
\begin{equation}
S_2=\frac1\beta\int_0^1d\tau\,\Big[g_{\mu\bar\nu}(x_0)\big(\dot q^\mu\dot{\bar q}^{\bar\nu}
+b^\mu\bar c^{\bar\nu}\big)+\bar\psi^1_\mu(\de_\tau+i\phi)\psi^\mu_1+\bar\psi^2_\mu(\de_\tau+i\theta)\psi^\mu_2\Big]\;,
\end{equation}
and an interaction part $S_{\text{int}}$
\begin{equation}
\begin{split}
S_{\text{int}} &= \frac1\beta\int_0^1d\tau\,\Big[\big(g_{\mu\bar\nu}(x_0+q)-g_{\mu\bar\nu}(x_0)\big)\big(\dot q^\mu\dot{\bar q}^{\bar\nu}
+b^\mu\bar c^{\bar\nu}\big)+\dot q^\nu\Gamma^\mu_{\nu\lambda}(x_0+q)\bar\psi^i_\mu\psi_i^\lambda\\
&-\frac{1}{2}R_\mu{}^\nu{}_\lambda{}^\sigma(x_0+q)\,\psi^\mu_i\bar\psi_\nu^i\psi^\lambda_j\bar\psi_\sigma^j \Big]\;.
\end{split}
\end{equation}
We denote as $\media{\,\bullet\,}$ the quantum average over the quadratic action
$$
\media{\,\bullet\,}=\frac{\int DqD\bar qDbD\bar cD\bar\psi D\psi\,\bullet\ {\rm e}^{-S_2}}{\int DqD\bar qDbD\bar cD\bar\psi D\psi\, {\rm e}^{-S_2}}\;.
$$
The partition function \eqref{partition function} finally reads
\begin{equation}\label{ea}
Z_\A[g]\propto\int_0^\infty\frac{d\beta}{\beta}\int_0^{2\pi}\frac{d\phi}{2\pi}\int_0^{2\pi}\frac{d\theta}{2\pi}\,\mu_\A(\phi,\theta)\int\frac{d^dx_0d^d\bar x_0}{(2\pi\beta)^d}g(x_0)\media{{\rm e}^{-S_{\text{int}}}}\;,
\end{equation}
where $(2\pi\beta)^{-d}$ is the usual free bosonic path integral, and we have plugged the Chern-Simons term as well as the free fermionic path integral in the modular measure
\begin{equation}
\mu_\A(\phi, \theta)=\tilde\mu_\A(\phi, \theta)\,{\rm e}^{-\tilde S_\A^{CS}}\,\left(2\cos\frac\phi2\right)^d\left(2\cos\frac\theta2\right)^d\;.
\end{equation}

Looking at \eqref{ea}, we see that the expression we obtained is quite compact: indeed, the perturbative computation of $\media{{\rm e}^{-S_{\text{int}}}}$ is common to all the four models we are interested in, and the choice of the model is entirely encoded in the form of the modular factor $\mu_\A(\phi, \theta)$, that is time to make more explicit.
Let us consider the $A$ theory: it gauges the supersymmetries and only the $U(1)\times U(1)$ subgroup. Since the $(p,q)$-form comes from a $(p+1,q+1)$ field strength, $F_{(p+1,q+1)}=\de\bar\de A_{(p,q)}$, according to \eqref{JiiA} the Chern-Simons couplings are given by $s_1=p+1-d/2$ and $s_2=d/2-q-1$, and the corresponding measure reads
\begin{equation}\label{mu A}
\mu_A(\phi, \theta)={\rm e}^{-i(p+1-d/2)\phi}{\rm e}^{i(q+1-d/2)\theta}\,\left(2\cos\frac\phi2\right)^{d-2}\left(2\cos\frac\theta2\right)^{d-2}\;.
\end{equation}
On the other hand, the $\alpha$ model has a single Chern-Simons coupling (see \eqref{Jiialpha}), but gauges the whole $U(2)$ group, giving the measure
\begin{equation}\label{mu alpha}
\mu_\alpha(\phi, \theta)=\frac12\,{\rm e}^{-i(p+1-d/2)(\phi+\theta)}\,\left(2\cos\frac\phi2\right)^{d-2}\left(2\cos\frac\theta2\right)^{d-2}\left(2\sin\frac{\phi-\theta}{2}\right)^2\;.
\end{equation}
In the remaining cases, where the supersymmetries are not gauged, the forms do not come from a field strength, and one has shifted Chern-Simons couplings: $s_1=p-d/2$, $s_2=d/2-q$. The resulting modular factors have the following form:
\begin{eqnarray}\label{mu B and beta}
\mu_B(\phi, \theta) &=& {\rm e}^{-i(p-d/2)\phi}{\rm e}^{i(q-d/2)\theta}\,\left(2\cos\frac\phi2\right)^{d}\left(2\cos\frac\theta2\right)^{d}\;,\\
\mu_\beta(\phi, \theta) &=& \frac12\,{\rm e}^{-i(p-d/2)(\phi+\theta)}\,\left(2\cos\frac\phi2\right)^{d}\left(2\cos\frac\theta2\right)^{d}\left(2\sin\frac{\phi-\theta}{2}\right)^2\;.
\end{eqnarray}

We are now ready to compute $\media{{\rm e}^{-S_{\text{int}}}}$ up to order $\beta^2$. Since we are free to use any coordinate system, we choose to employ K\"ahler normal coordinates (see \cite{Higashijima:2000wz}, for instance) centered at $x_0$, that allow to maintain explicit covariance under reparametrization of $x_0$ while keeping holomorphic coordinates at each step.
Denoting with $S_n$ the part of $S_{\text{int}}$ containing $n$-fields vertices, it turns out that the only terms giving non vanishing contribution up to order $\beta^2$ are the following ones
\begin{equation}\label{S_4 and S_6}
\begin{split}
S_4 &= \frac1\beta\int_0^1 d\tau\,\Big[R_{\mu\bar\nu\lambda\bar\sigma}\,q^\lambda\bar q^{\bar\sigma}\big(\dot q^\mu\dot{\bar q}^{\bar\nu}
+b^\mu\bar c^{\bar\nu}\big)-R^\lambda{}_{\sigma\bar\nu\mu}\,\dot q^\mu\bar q^{\bar\nu}\psi_i^\sigma\bar\psi^i_\lambda-\frac{1}{2}R_\mu{}^\nu{}_\lambda{}^\sigma\,\psi^\mu_i\bar\psi_\nu^i\psi^\lambda_j\bar\psi_\sigma^j\Big]\\
S_6 &= \frac1\beta\int_0^1 d\tau\,\Big[\frac14\big[\nabla_{(\bar\sigma}\nabla_\lambda R_{\mu\bar\nu\rho\bar\kappa)}
+3R^{\bar\tau}{}_{(\bar\nu\lambda\bar\kappa}R_{\mu\bar\sigma\rho)\bar\tau}\big]q^\lambda\bar q^{\bar\sigma}q^\rho\bar q^{\bar\kappa}
\big(\dot q^\mu\dot{\bar q}^{\bar\nu}+b^\mu\bar c^{\bar\nu}\big)\\
&-\frac12\big[\nabla_\rho\nabla_{\bar\sigma}R^\lambda{}_{\mu\bar\lambda\nu}+R^{\bar\tau}{}_{\bar\lambda\rho\bar\sigma}
R^\lambda{}_{\mu\bar\tau\nu}\big]q^\rho\bar q^{\bar\sigma}\bar q^{\bar\lambda}\dot q^\mu\psi_i^\nu\bar\psi^i_\lambda-\frac{1}{2}\nabla_\rho\nabla_{\bar\tau}R_\mu{}^\nu{}_\lambda{}^\sigma\,q^\rho\bar q^{\bar\tau}\psi^\mu_i\bar\psi_\nu^i\psi^\lambda_j\bar\psi_\sigma^j\Big]\;,
\end{split}
\end{equation}
where all tensors are calculated at $x_0$ and round brackets denote weighted symmetrization, separately among holomorphic and anti-holomorphic indices.

The two-point functions are readily computed from the free action, and are given by
\begin{equation}\label{2 point functions}
\begin{split}
\media{q^{\mu}(\tau)\bar q^{\bar\nu}(\sigma)} &= -\beta g^{\mu\bar\nu}(x_0)\Delta(\tau,\sigma)\;,
\quad\media{b^\mu(\tau)\bar c^{\bar\nu}(\sigma)}=-\beta g^{\mu\bar\nu}(x_0)\delta(\tau,\sigma) \\[1mm]
\media{\psi_i^\mu(\tau)\bar\psi^j_\nu(\sigma)} &= \beta\, \delta^j_i\,\delta^\mu_\nu\Delta_{\text{f}}(\tau-\sigma,\phi_i)
\end{split}
\end{equation}
where $\phi_i\equiv(\phi, \theta)$, and the propagators in the continuum limit read
\begin{equation}\label{propagators}
\begin{split}
\Delta(\tau,\sigma) &= \sigma(\tau-1)\,\theta(\tau-\sigma)+\tau(\sigma-1)\,\theta(\sigma-\tau)\;,\\
\Delta_{\text{f}}(\tau-\sigma,\phi_i) &= \frac{{\rm e}^{-i\phi_i (\tau-\sigma)}}{2\cos\frac{\phi_i}{2}}\big[{\rm e}^{i\frac{\phi_i}{2}}\theta(\tau-\sigma)-{\rm e}^{-i\frac{\phi_i}{2}}\theta(\sigma-\tau)\big]
\end{split}
\end{equation}
with $\theta(\tau-\sigma)$ the Heaviside step function and  $\delta(\tau,\sigma)$ the Dirac delta acting on functions that vanish at the endpoints.
It is well known that path integrals in curved space require regularization. Indeed one can see from \eqref{propagators} that in the computations one
has to face products and derivatives of such distributions, that generically are ill defined. Here we choose to employ Time Slicing (TS) regularization \cite{Bastianelli:2006rx, DeBoer:1995hv, deBoer:1995cb}, that gives unambiguous prescriptions on how to handle these subtleties, and does not require counterterms (the standard TS counterterm of the $N=2$ sigma model vanishes on K\"ahler manifolds).
Among the usual regularization schemes the TS rules are the simpler\footnote{
Other known regularizations are Mode Regularization (MR) \cite{Bastianelli:1998jm, Bastianelli:1998jb, Bonezzi:2008gs}
and Dimensional Regularization (DR) \cite{Kleinert:1999aq,Bastianelli:2000pt,Bastianelli:2000nm}.
MR carries a non covariant counterterm that is non vanishing on K\"ahler manifolds, so that additional vertices must be included to obtain the correct final answer.
DR is covariant, but requires a few integration by parts for its implementation, so that we found TS to be the simplest one for the present calculations.
All of these regularizations have been recently extended to nonlinear sigma models with $N$ supersymmetries in \cite{Bastianelli:2011cc}.
}:when computing the various Feynman diagrams the delta functions
have to be treated as Kronecker deltas, and the Heaviside theta has the regulated value $\theta(0)=\frac12$. Moreover, the ghost system forbids the appearence of products of delta functions. By means of these prescriptions, the propagators \eqref{propagators} and their derivatives have well defined equal-time expressions, that are listed for convenience in Appendix \ref{app:propagators}

From \eqref{S_4 and S_6}-\eqref{2 point functions} we see that each piece $S_n$ of $S_{\text{int}}$ gives a contribution
of order $\beta^{n/2-1}$. Therefore, our quantum average can be written explicitly as
\begin{equation}
\media{{\rm e}^{-S_{\text{int}}}}=1-\media{S_4}-\media{S_6}+\frac12\media{S_4^2}+{\cal O}(\beta^3)\;.
\end{equation}
Using the expressions given in \eqref{S_4 and S_6} and TS prescriptions in calculating Feynman diagrams, one eventually obtains
\begin{equation}\label{master formula}
\begin{split}
&\media{{\rm e}^{-S_{\text{int}}}} = 1+\beta\Big(-\frac{1}{12}+\frac14\tan\frac\phi2\tan\frac\theta2\Big)\,R\\
&+\beta^2\,\Big\{\Big[\frac{1}{180}
-\frac{1}{96}\Big(\cos^{-2}\frac\phi2+\cos^{-2}\frac\theta2\Big)+\frac{1}{32}\cos^{-2}\frac\phi2\cos^{-2}\frac\theta2\Big]\,R_{\mu\bar\nu\lambda\bar\sigma}R^{\mu\bar\nu\lambda\bar\sigma}\\
&+\Big[-\frac{17}{720}+\frac{1}{24}\Big(\cos^{-2}\frac\phi2+\cos^{-2}\frac\theta2\Big)-\frac{1}{16}\cos^{-2}\frac\phi2\cos^{-2}\frac\theta2
+\frac{1}{48}\tan\frac\phi2\tan\frac\theta2\Big]\,R_{\mu\bar\nu}R^{\mu\bar\nu}\\
&+\Big[\frac{5}{144}-\frac{1}{32}\Big(\cos^{-2}\frac\phi2+\cos^{-2}\frac\theta2\Big)+\frac{1}{32}\cos^{-2}\frac\phi2\cos^{-2}\frac\theta2
-\frac{1}{48}\tan\frac\phi2\tan\frac\theta2\Big]\,R^2\\&
+\Big[-\frac{1}{240}+\frac{1}{48}\tan\frac\phi2\tan\frac\theta2\Big]\, \nabla^2R\Big\}\;,
\end{split}
\end{equation}
where $\nabla^2R=2g^{\mu\bar\nu}\de_\mu\de_{\bar\nu}R$.
This is our perturbative master formula, from which we can easily compute the first Seeley-DeWitt coefficients
(SDW or heat kernel coefficients)
for our different models by plugging \eqref{master formula} into \eqref{partition function}, and performing the modular integrals over $\phi$ and $\theta$ with the different measures $\mu_\A(\phi, \theta)$. At this stage a first subtlety arises in performing such integrations: by looking at the measures $\mu_\A$ and at the expansion \eqref{master formula}, one sees that, given the complex dimension $d$, at sufficiently large orders in $\beta$ two poles appear along the integration paths, namely at $\phi=\pi$ and $\theta=\pi$. This corresponds to effectively giving periodic boundary conditions to fermionic fields that develop zero modes one has to deal with
(see \emph{e.g.} \cite{Bastianelli:2005vk}).
In order to find and understand the correct prescription, it is useful to switch to Wilson loop variables, \emph{i.e.} $z={\rm e}^{i\phi}$ and $w={\rm e}^{i\theta}$. The modular integrals then turn into contour integrals on the unit circle centered in zero in the complex $z$ and $w$ planes, that we denote by $\gamma$
$$
\int_0^{2\pi}\frac{d\phi}{2\pi}\int_0^{2\pi}\frac{d\theta}{2\pi}=\oint_\gamma\frac{dz}{2\pi iz}\oint_\gamma\frac{dw}{2\pi iw}\;.
$$
The possible poles now show up on the integration contour at $z=-1$ or $w=-1$. To correctly deal with such poles, it turns out that one has to slightly deform the two contours excluding the pole in $z=-1$,
with the regulated contour $\gamma^-$ shown in figure 1,
and including the pole in $w=-1$ with the regulated contour $\gamma^+$ shown in figure 2.
\begin{figure}[h]
  \centering
  \includegraphics[width=2.5in]{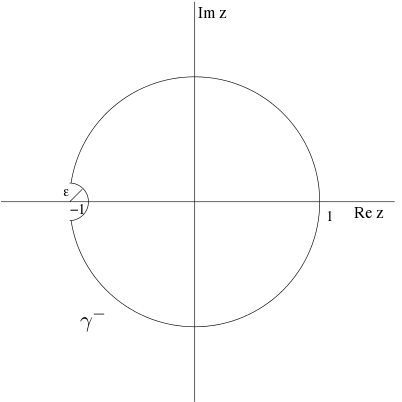}
  \caption[Fig2]%
  {The regulated contour $\gamma^-$ that excludes the pole at $z=-1$.}
\end{figure}
The heuristic reason for this choice is that the correct known results, for instance the SDW coefficients for a scalar field, only come out from the aforementioned prescription. Nevertheless, we can justify the choice of two different contours for the $z$ and $w$ integrals: it corresponds to using two different bases for the two fermionic species. While the first one is realized treating $\psi_1$'s as creation and $\bar\psi^1$'s as annihilation operators, the second has $\bar\psi^2$'s as creators and $\psi_2$'s as annihilators. The first realization leads to the ``standard'' contour $\gamma^-$ for $z$ (see \cite{Bastianelli:2011pe}), while the second, that is obtained from the usual one with an anti-holomorphic dualization of the fermionic vacuum\footnote{Denoting with $\ket{0}_{\bar\psi^2}$ the Fock vacuum annihilated by the $\bar\psi^2$ operators, and with $\ket{0}_{\psi_2}$ the Fock vacuum annihilated by the $\psi_2$'s, one has $\ket{0}_{\psi_2}\propto e\,\epsilon_{\mu_1.!
 ..\mu_d}\,\psi^{\mu_1}_2...\psi^{\mu_d}_2\ket{0}_{\bar\psi^2}$. Note that in the main text we use bra coherent states, dual to ket coherent states built from $\ket{0}_{\psi_2}$, so that throughout the paper the $\bar \psi ^2$'s indicate the eigenvalues (Grassmann numbers) of the corresponding operators, and no confusion should arise.}, leads to the $\gamma^+$ contour for $w$. Hence, at the level of differential forms, switching from $\gamma^-$ to $\gamma^+$ contour corresponds to performing an (anti)-holomorphic Hodge duality\footnote{The total Hodge duality involves the full $\epsilon$ tensor $g\,\epsilon_{\mu_1...\mu_d\bar\nu_1...\bar\nu_d}$, while the holomorphic one requires only $e\,\epsilon_{\mu_1...\mu_d}$}. As shown in \cite{Bastianelli:2011pe}, this introduces an extra coupling of the fields to the $U(1)$ part of the K\"ahler connection, that is not allowed in the present models.

\begin{figure}[h]
  \centering
  \includegraphics[width=2.5in]{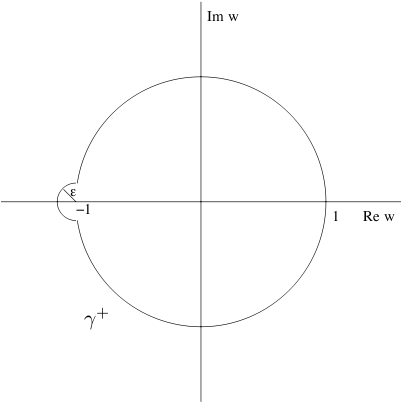}
  \caption[Fig2]%
  {The regulated contour $\gamma^+$ that includes the pole at $w=-1$.}
\end{figure}

Given the correct contour prescriptions, we are now ready to compute the SDW coefficients for the differential forms and study their duality properties.
We shall define the SDW coefficients in the expansion of the partition functions as:
\begin{equation}\label{Seeley coefficients parametrized}
Z\hspace{-1mm}\propto\hspace{-1mm}\int_0^\infty\hspace{-1mm}\frac{d\beta}{\beta}\hspace{-1mm}\int\frac{d^dx_0d^d\bar x_0}{(2\pi\beta)^d}
g(x_0)\Big\{v_1+v_2\beta\,R+\beta^2\Big[v_3\,
R_{\mu\bar\nu\lambda\bar\sigma}R^{\mu\bar\nu\lambda\bar\sigma}
+v_4\,R_{\mu\bar\nu}R^{\mu\bar\nu}+v_5\,R^2+v_6\,\nabla^2R\Big]\Big\}.
\end{equation}
Let us notice that in standard quantum field theories the first coefficient $v_1$ gives the number of physical degrees of freedom, and vanishes when one treats non-propagating fields.
A similar interpretation is applicable in our case, where a K\"ahler manifold takes the role of spacetime.
We will list the coefficients for a generic form $\A$ as: $\A\to(v_1;v_2;v_3;v_4;v_5;v_6)$

Let us consider first the $A$ model of $(p,q)$-form gauge fields. Plugging the master formula \eqref{master formula} in \eqref{partition function} with the measure \eqref{mu A} we have, using the Wilson loop variables,
\begin{equation}\label{Z A}
Z_A[g]\propto\int_0^\infty\frac{d\beta}{\beta}\oint_{\gamma^-}\frac{dz}{2\pi iz}\oint_{\gamma^+}\frac{dw}{2\pi iw}\frac{1}{z^p\,w^{d-2-q}}\big[(z+1)(w+1)\big]^{d-2}\!\int\frac{d^dx_0d^d\bar x_0}{(2\pi\beta)^d}g(x_0)\media{{\rm e}^{-S_{\text{int}}}}\;.
\end{equation}
Upon performing the modular integrations one gets the following SDW coefficients
\begin{equation}\label{A form coefficients}
\begin{split}
&A_{(p,q)}\to  \binom{d-2}{p}\binom{d-2}{q}\times\left(1;\,\frac16-\frac{p(d-2-q)+q(d-2-p)}{2(d-2)^2};\right.\\[3mm]
&\frac{1}{180}-\frac{p(d-2-p)+q(d-2-q)}{24(d-2)(d-3)}+\frac{p(d-2-p)q(d-2-q)}{2(d-2)^2(d-3)^2};\\[3mm]
&-\frac{1}{360}+\frac{p(d-2-p)+q(d-2-q)}{6(d-2)(d-3)}-\frac{p(d-2-p)q(d-2-q)}{(d-2)^2(d-3)^2}-\frac{p(d-2-q)+q(d-2-p)}{24(d-2)^2};\\[3mm]
&\frac{1}{72}-\frac{p(d-2-p)+q(d-2-q)}{8(d-2)(d-3)}+\frac{p(d-2-p)q(d-2-q)}{2(d-2)^2(d-3)^2}+\frac{p(d-2-q)+q(d-2-p)}{24(d-2)^2};\\[3mm]
&\left.\frac{1}{60}-\frac{p(d-2-q)+q(d-2-p)}{24(d-2)^2}\right)\;.
\end{split}
\end{equation}
This formula is valid for $0\leq p,q\leq d-2$, since for larger $p$ or $q$ the field does not propagate, and for $d\geq4$, since in lower dimensions the pole in $z,w=-1$ appears also in these first coefficients, and one has to use the regulated contours.
We can see that \eqref{A form coefficients} correctly reproduces the coefficients for a scalar field, obtained by setting $p=q=0$
\begin{equation}\label{scalar}
A_{(0,0)}\to\left(1;\,\frac16;\,\frac{1}{180};\,-\frac{1}{360};\,\frac{1}{72};\,\frac{1}{60}\right)\;,
\end{equation}
as well as for $(p,0)$-forms, as we can compare \eqref{A form coefficients} for $A_{(p,0)}$ with the results of \cite{Bastianelli:2011pe} with the charge set to zero\footnote{For $q=0$ or $p=0$ the compensator term drops
out from the field equation of our $A_{(p,q)}$ forms, that reduce to the $(p,0)$-forms studied in \cite{Bastianelli:2011pe} with zero charge.}
\begin{equation}
\begin{split}
A_{(p,0)}\to  &\binom{d-2}{p}\times\left(1;\,\frac16-\frac{p}{2(d-2)};\,\frac{1}{180}-\frac{p(d-2-p)}{24(d-2)(d-3)};\right.\\[3mm]
&-\frac{1}{360}+\frac{p(d-2-p)}{6(d-2)(d-3)}-\frac{p}{24(d-2)};\,\frac{1}{72}-\frac{p(d-2-p)}{8(d-2)(d-3)}+\frac{p}{24(d-2)};\\[3mm]
&\left.\frac{1}{60}-\frac{p}{24(d-2)}\right)\;.
\end{split}
\end{equation}
By inspecting the coefficients \eqref{A form coefficients}, it is manifest the symmetry under the exchange $p\leftrightarrow q$ that corresponds to complex conjugation: $A_{(p,q)}\sim A_{(q,p)}$, and under $p\leftrightarrow d-2-q$, that corresponds to Hodge duality $A_{(p,q)}\sim A_{(d-2-q,d-2-p)}$. We will see in the following that the first symmetry is fully preserved in the effective action, while the Hodge duality suffers a topological mismatch that will be investigated at the non-perturbative level in the next section.

We now turn to computing the SDW coefficients in lower dimensions. In three complex dimensions propagating fields have $p$ and $q$ restricted to be zero or one, and we get
\begin{equation}\label{A form coeff d=3}
\begin{split}
&d=3\;,\quad p,q=\{0,1\}\\[2mm]
&A_{(p,q)}\to\left(1;\,\frac16-\frac{p(1-q)+q(1-p)}{2};\,\frac{1}{180}-\frac{p+q}{24}+\frac12\,pq;\right.\\[2mm]
&\hspace{20mm}-\frac{1}{360}+\frac{p+q}{6}-pq
-\frac{p(1-q)+q(1-p)}{24};\\[2mm]
&\left.\hspace{20mm}\frac{1}{72}-\frac{p+q}{8}+\frac12\,pq
+\frac{p(1-q)+q(1-p)}{24};\,\frac{1}{60}-\frac{p(1-q)+q(1-p)}{24}\right)\;.
\end{split}
\end{equation}
It can be successfully compared with the results of \cite{Bastianelli:2011pe} for the scalar and the $(1,0)$ form setting the charge appearing there to zero. The formula \eqref{A form coeff d=3} is still invariant under the exchange of $p$ and $q$, that tells $A_{(1,0)}\sim A_{(0,1)}$, but the symmetry under Hodge duality $p\to1-q$, $q\to1-p$ is lost, due to topological mismatches, and indeed one has
\begin{equation}
\begin{split}
&d=3\\[2mm]
&A_{(0,0)}\to\left(1;\,\frac16;\,\frac{1}{180};\,-\frac{1}{360};\,\frac{1}{72};\,\frac{1}{60}\right)\;,\\[2mm]
&A_{(1,1)}\to\left(1;\,\frac16;\,\frac{19}{45};\,-\frac{241}{360};\,\frac{19}{72};\,\frac{1}{60}\right)\;.
\end{split}
\end{equation}
In $d=2$ only the scalar propagates, and by means of \eqref{Z A} one finds the correct results already presented
\begin{equation}
d=2\;,\quad A_{(0,0)}\to\left(1;\,\frac16;\,\frac{1}{180};\,-\frac{1}{360};\,\frac{1}{72};\,\frac{1}{60}\right)\;.
\end{equation}
We are not interested in $d=1$ in the present paper, since the field equations are somewhat degenerate from the very beginning, see \emph{e.g.} \cite{Bastianelli:2011pe}.

Now we can turn to the $\alpha$ model that describes $(p,d-2-p)$-form gauge fields. It is worth stressing that the present theory is definitely not obtainable by setting $q=d-2-p$ in the $A_{(p,q)}$ forms, since they obey different field equations. Indeed, by means of the measure \eqref{mu alpha}, the partition function is given by
\begin{equation}\label{Z alpha}
\begin{split}
Z_\alpha[g]\propto & -\frac12\int_0^\infty\frac{d\beta}{\beta}\oint_{\gamma^-}\frac{dz}{2\pi iz}\oint_{\gamma^+}\frac{dw}{2\pi iw}\frac{1}{(z\,w)^{p+1}}\big[(z+1)(w+1)\big]^{d-2}(z-w)^2\\
&\times\int\frac{d^dx_0d^d\bar x_0}{(2\pi\beta)^d}g(x_0)\media{{\rm e}^{-S_{\text{int}}}}\;.
\end{split}
\end{equation}
The corresponding Seeley-DeWitt coefficients are given by
\begin{equation}\label{alpha form coefficients}
\begin{split}
&\alpha_{(p,d-2-p)}\to\binom{d-2}{p}\frac{(d-1)!}{(p+1)!(d-1-p)!}\times\left(1;\,-\frac13+\frac{p(d-2-p)}{(d-1)(d-2)};\right.\\[3mm]
&\frac{1}{180}+\frac{p(d-2-p)}{12(d-1)(d-2)^2(d-3)}[6(p+1)(d-1-p)-d(d-2)];\\[3mm]
&-\frac{2}{45}+\frac{p(d-2-p)}{12(d-1)(d-2)^2(d-3)}[(d-2)(5d-3)-12(p+1)(d-1-p)];\\[3mm]
&\frac{1}{18}+\frac{p(d-2-p)}{12(d-1)(d-2)^2(d-3)}[6(p+1)(d-1-p)-(d-2)(4d-3)];\\[3mm]
&\left.-\frac{1}{40}+\frac{p(d-2-p)}{12(d-1)(d-2)}\right)\;.
\end{split}
\end{equation}
We notice that \eqref{alpha form coefficients} is explicitly invariant under $p\leftrightarrow d-2-p$, corresponding to complex conjugation: $\alpha_{(p,d-2-p)}\sim\alpha_{(d-2-p,p)}$. In this model the Hodge duality acts trivially, since it sends $\alpha_{(p,d-2-p)}$ into itself and indeed no topological mismatch can ever arise in this context, as we will show in more detail in the next section.
As in the previous case this formula is valid in $d\geq4$ and for $0\leq p\leq d-2$, and the lower dimensional cases will be presented separately. We notice that this model never describes a scalar field, except in two complex dimensions. The only check we can do is for $p=0,d-2$, since in this particular case the $\alpha_{(d-2,0)}$ form should be the same as the $A_{(d-2,0)}$. In $d\geq4$ the check is indeed successful, and matches also the result of \cite{Bastianelli:2011pe}:
\begin{equation}
d\geq4\;,\quad\alpha_{(d-2,0)}\to\left(1;\,-\frac13;\,\frac{1}{180};\,-\frac{2}{45};\,\frac{1}{18};\,-\frac{1}{40}\right)\;,
\end{equation}
however, this point contains a nontrivial subtlety.
From Sec. \ref{sec flat space} one can see that the difference between the field equations for $\alpha$ and $A$ forms is essentially in the trace term. Therefore, it is natural to think that, at the level of effective actions, one should be able to obtain the $\alpha$ contribution from the $A$ model by subtracting some trace terms. This is indeed the case, as we will prove in sec. \ref{subsec anatomy}, and one has the following exact relation
\begin{equation}
Z^\alpha_{p,d-2-p}=Z^A_{p,d-2-p}-\frac12\,Z^A_{p-1,d-3-p}-\frac12\,Z^A_{p+1,d-1-p}\;,
\end{equation}
where we see that the subtractions involve the trace and its dual. The subtlety arises for $p=0$ or $p=d-2$: in this case, one would naively expect the subtraction terms to be zero. While this is true for negative degree forms, the term $Z^A_{1,d-1}$ is nonzero, although non-propagating and purely topological, and the effective action becomes:
\begin{equation}\label{alpha mismatch}
Z^\alpha_{0,d-2}=Z^A_{0,d-2}-\frac12\,Z^{A\,\text{top}}_{1,d-1}\;,
\end{equation}
where we stressed the topological nature of $A_{(1,d-1)}$.

This topological contribution is not visible in $d\geq4$ at order $\beta^2$, but one can find it in lower dimensions. For instance, in three complex dimensions the only physical $\alpha$ field is $\alpha_{(1,0)}\sim\alpha_{(0,1)}$, with SDW coefficients given by
\begin{equation}
\begin{split}
&d=3\\[2mm]
&\alpha_{(0,1)}\to\left(1;\,-\frac13;\,\frac{139}{720};\,-\frac{53}{180};\,\frac{17}{144};\,-\frac{1}{40}\right)\;,
\end{split}
\end{equation}
that should naively equal
\begin{equation}
\begin{split}
&d=3\\[2mm]
&A_{(0,1)}\to\left(1;\,-\frac13;\,-\frac{13}{360};\,\frac{11}{90};\,-\frac{5}{72};\,-\frac{1}{40}\right)\;,
\end{split}
\end{equation}
given by \eqref{A form coeff d=3}. The mismatch is due to the topological $A_{(1,2)}$ form, whose contribution can be computed from \eqref{Z A} and reads
\begin{equation}
\begin{split}
&d=3\\[2mm]
&A_{(1,2)}\to\left(0;\,0;\,-\frac{11}{24};\,\frac{5}{6};\,-\frac{3}{8};\,0\right)\;,
\end{split}
\end{equation}
indeed satisfying \eqref{alpha mismatch}, that tells
\begin{equation}
Z^\alpha_{0,1}=Z^A_{0,1}-\frac12\,Z^{A\,\text{top}}_{1,2}\;.
\end{equation}
The other check we can perform is in $d=2$, where the only physical field should be a scalar $\alpha_{(0,0)}$. Its coefficients are
\begin{equation}
\begin{split}
&d=2\\[2mm]
&\alpha_{(0,0)}\to\left(1;\,-\frac13;\,-\frac{11}{45};\,\frac{41}{90};\,-\frac{7}{36};\,-\frac{1}{40}\right)\;,
\end{split}
\end{equation}
that differ from the usual scalar ones \eqref{scalar} by the topological contribution of $A_{(1,1)}$
\begin{equation}
\begin{split}
&d=2\\[2mm]
&A_{(1,1)}\to\left(0;\,1;\,\frac{1}{2};\,-\frac{11}{12};\,\frac{5}{12};\,\frac{1}{12}\right)\;,
\end{split}
\end{equation}
and satisfy
\begin{equation}
Z^\alpha_{0,0}=Z^A_{0,0}-\frac12\,Z^{A\,\text{top}}_{1,1}\;.
\end{equation}

We concentrate now on the remaining models, \emph{i.e.} non gauge differential forms. In fact, the $B$ and $\beta$ forms do not enjoy any gauge symmetry, and obey simple wave equations given by the Laplace operator acting on forms.
Let us start with the $B$ model containing non gauge $(p,q)$-forms $B_{(p,q)}$. The first measure of \eqref{mu B and beta} allows us to write the partition function as:
\begin{equation}\label{Z B}
Z_B[g]\propto\int_0^\infty\frac{d\beta}{\beta}\oint_{\gamma}\frac{dz}{2\pi iz}\oint_{\gamma}\frac{dw}{2\pi iw}\frac{1}{z^p\,w^{d-q}}\big[(z+1)(w+1)\big]^{d}\!\int\frac{d^dx_0d^d\bar x_0}{(2\pi\beta)^d}g(x_0)\media{{\rm e}^{-S_{\text{int}}}}\;.
\end{equation}
We denoted the two contours as $\gamma$ since, as we will show in the following, the absence of the ghost terms for the supersymmetry avoids the presence of poles along the contour at the non-perturbative level.
The effective action of these non gauge $(p,q)$-forms is characterized by the following coefficients, that we compute from \eqref{Z B}
\begin{equation}\label{B form coefficients}
\begin{split}
&B_{(p,q)}\to  \binom{d}{p}\binom{d}{q}\times\left(1;\,\frac16-\frac{p(d-q)+q(d-p)}{2d^2};\right.\\[3mm]
&\frac{1}{180}-\frac{p(d-p)+q(d-q)}{24d(d-1)}+\frac{p(d-p)q(d-q)}{2d^2(d-1)^2};\\[3mm]
&-\frac{1}{360}+\frac{p(d-p)+q(d-q)}{6d(d-1)}-\frac{p(d-p)q(d-q)}{d^2(d-1)^2}-\frac{p(d-q)+q(d-p)}{24d^2};\\[3mm]
&\frac{1}{72}-\frac{p(d-p)+q(d-q)}{8d(d-1)}+\frac{p(d-p)q(d-q)}{2d^2(d-1)^2}+\frac{p(d-q)+q(d-p)}{24d^2};\\[3mm]
&\left.\frac{1}{60}-\frac{p(d-q)+q(d-p)}{24d^2}\right)\;.
\end{split}
\end{equation}
This result is valid for any dimension we are interested in. It is in fact singular only in $d=1$, that we do not wish to consider. The effective action is manifestly invariant under complex conjugation $p\leftrightarrow q$ and Hodge duality $p\leftrightarrow d-q$, and no one of these symmetries is affected by topological issues, as it will be proved soon. For $q=0$ \eqref{B form coefficients} reproduces the result found in \cite{Bastianelli:2011pe}, including the scalar field recovered by setting also $p=0$
\begin{equation}
\begin{split}
B_{(p,0)}\to  &\binom{d}{p}\times\left(1;\,\frac16-\frac{p}{2d};\,\frac{1}{180}-\frac{p(d-p)}{24d(d-1)};\,-\frac{1}{360}+\frac{p(d-p)}{6d(d-1)}-\frac{p}{24d};\right.\\[3mm]
&\left.\frac{1}{72}-\frac{p(d-p)}{8d(d-1)}+\frac{p}{24d};\,\frac{1}{60}-\frac{p}{24d}\right)\;,\\[3mm]
B_{(0,0)}\to  &\left(1;\,\frac16;\,\frac{1}{180};\,-\frac{1}{360};\,\frac{1}{72};\,\frac{1}{60}\right)\;.
\end{split}
\end{equation}

The last model we are interested in describes $(p,d-p)$ traceless forms with no gauge symmetry. Their partition function is recovered using the second modular measure of \eqref{mu B and beta} and yields
\begin{equation}\label{Z beta}
\begin{split}
Z_\beta[g]\propto & -\frac12\int_0^\infty\frac{d\beta}{\beta}\oint_{\gamma}\frac{dz}{2\pi iz}\oint_{\gamma}\frac{dw}{2\pi iw}\frac{1}{(z\,w)^{p+1}}\big[(z+1)(w+1)\big]^{d}(z-w)^2\\
&\times\int\frac{d^dx_0d^d\bar x_0}{(2\pi\beta)^d}g(x_0)\media{{\rm e}^{-S_{\text{int}}}}\;.
\end{split}
\end{equation}
The Seeley-DeWitt coefficients are readily obtained from \eqref{master formula} and \eqref{Z beta} as
\begin{equation}\label{beta form coefficients}
\begin{split}
&\beta_{(p,d-p)}\to\binom{d}{p}\frac{(d+1)!}{(p+1)!(d+1-p)!}\times\left(1;\,-\frac13+\frac{p(d-p)}{(d+1)d};\right.\\[3mm]
&\frac{1}{180}+\frac{p(d-p)}{12d^2(d+1)(d-1)}[6(p+1)(d+1-p)-d(d+2)];\\[3mm]
&-\frac{2}{45}+\frac{p(d-p)}{12d^2(d+1)(d-1)}[d(5d+7)-12(p+1)(d+1-p)];\\[3mm]
&\frac{1}{18}+\frac{p(d-p)}{12d^2(d+1)(d-1)}[6(p+1)(d+1-p)-d(4d+5)];\\[3mm]
&\left.-\frac{1}{40}+\frac{p(d-p)}{12d(d+1)}\right)\;,
\end{split}
\end{equation}
that are valid in $d>1$ and exhibit explicit symmetry under complex conjugation: $p\to d-p$, whereas Hodge duality acts trivially as in the $\alpha$ theory.

\section{Dualities}\label{sec dualities}

In this section we will study in more depth exact duality relations and the topological mismatches that potentially arise in this context. Before diving into the duality issues, it is useful to study at the non-perturbative level the structure of our effective actions and the relations between the different models.

\subsection{Anatomy of the effective actions}\label{subsec anatomy}

In order to uncover exact relations between the effective actions, it is fruitful to switch to a mixed formalism, in between the functional and the operatorial ones. Since the hamiltonian $H$ and the fermion number operators contained in $J_1^1$ and $J_2^2$ commute with each other, they are simultaneously diagonalizable, and the Hilbert space can be viewed as a direct sum of subspaces with definite fermion numbers $N=n$, $\bar N=m$: $\mathcal{H}=\bigoplus_{n,m=0}^d\mathcal{H}_{n,m}$. Thus, we decide to leave the modular integrals as they stand, and to recast the rest of the path integral as a quantum mechanical trace over the Hilbert space. To display this, let us focus on the $A$ model first. We rewrite the partition function \eqref{partition function} as
\begin{equation}\label{Z structure explained}
\begin{split}
Z^A_{p,q}&\propto\int_0^\infty\frac{d\beta}{\beta}\int_0^{2\pi}\frac{d\phi}{2\pi}\int_0^{2\pi}\frac{d\theta}{2\pi}\,\left(2\cos\frac\phi2\right)^{-2}\left(2\cos\frac\theta2\right)^{-2}\;\int_{\text{P}}\Dc x\Dc\bar x\int_{\text{A}}D\bar\psi D\psi\, {\rm e}^{-S_A[X,\tilde G_A]}\\[2mm]
&=\int_0^\infty\frac{d\beta}{\beta}\int_0^{2\pi}\frac{d\phi}{2\pi}\int_0^{2\pi}\frac{d\theta}{2\pi}\,
\left(2\cos\frac\phi2\right)^{-2}\left(2\cos\frac\theta2\right)^{-2}\,\Tr\Big[{\rm e}^{i\phi(J_1^1-s_1)}{\rm e}^{i\theta(J_2^2-s_2)}{\rm e}^{-\beta H}\Big]\\[2mm]
&=\int_0^\infty\frac{d\beta}{\beta}\underbrace{\oint_{\gamma^-}\frac{dz}{2\pi iz}\oint_{\gamma^+}\frac{dw}{2\pi iw}\,\frac{z}{(z+1)^2}\frac{w}{(w+1)^2}\,\Tr\Big[z^{(J_1^1-s_1)}w^{(J_2^2-s_2)}{\rm e}^{-\beta H}\Big]}_{\mathcal{Z}^A_{p,q}(\beta)}\;,
\end{split}
\end{equation}
where $s_1=p+1-d/2$ and $s_2=d/2-q-1$, and where we defined the partition function density in proper time $\mathcal{Z}(\beta)$. In this formula we have rewritten the path integral over matter fields as a quantum mechanical trace. In doing so we made explicit the dependence on the Wilson loop variables $z={\rm e}^{i\phi}$ and $w={\rm e}^{i\theta}$. We recall that the factors $\frac{z}{(z+1)^2}\frac{w}{(w+1)^2}$ are the contributions of the SUSY superghosts, and $H$ is the quantum hamiltonian given in section \ref{sec curved space}. As a last notational issue, as we have mentioned that the Hilbert space can be viewed as a sum of sectors with fixed fermion numbers,  we are going to denote by $t_{n,m}$ the trace of ${\rm e}^{-\beta H}$ taken over the Hilbert subspace with fixed $N=n$ and $\bar N=m$
\begin{equation}
t_{n,m}(\beta)\equiv\Tr_{\mathcal{H}_{n,m}}\Big[{\rm e}^{-\beta H}\Big]\;.
\end{equation}

Having presented our new notations, we start by analyzing the simplest of our models, namely the $B_{(p,q)}$ forms.

The $B$ model does not gauge supersymmetry, hence the factors coming from the superghosts are absent. Moreover, in this case we do not start from curvatures to arrive at the $B$ forms, and the Chern-Simons couplings read
$$
s_1=p-\frac{d}{2}\;,\quad s_2=\frac{d}{2}-q\;,
$$
so that
$$
J_1^1-s_1=N-p\;,\quad J^2_2-s_2=-\bar N+q\;.
$$
The partition function density $\mathcal{Z}^B_{p,q}(\beta)$ in our present notation reads
\begin{equation}\label{B form exact}
\begin{split}
\mathcal{Z}^B_{p,q}(\beta)&=\oint_{\gamma}\frac{dz}{2\pi iz}\oint_{\gamma}\frac{dw}{2\pi iw}\,\Tr\Big[z^{(J_1^1-s_1)}w^{(J_2^2-s_2)}{\rm e}^{-\beta H}\Big]\\
&=\oint_{\gamma}\frac{dz}{2\pi iz}\oint_{\gamma}\frac{dw}{2\pi iw}\,\Tr\Big[z^{(N-p)}w^{(-\bar N+q)}{\rm e}^{-\beta H}\Big]\\
&=\oint_{\gamma}\frac{dz}{2\pi iz}\oint_{\gamma}\frac{dw}{2\pi iw}\,\sum_{n,m=0}^dz^{n-p}w^{q-m}\,\Tr_{\mathcal{H}_{n,m}}\Big[{\rm e}^{-\beta H}\Big]=t_{p,q}(\beta)\;.
\end{split}
\end{equation}
We can see from \eqref{B form exact} that, not having gauged the supersymmetries, no pole can arise in $z=-1$ and $w=-1$ at the non-perturbative level, and this justifies the use of the unregulated contour $\gamma$. The result is given indeed by the single residue at $z=w=0$, and the effective action for $(p,q)$-forms obeying just a Laplace-type equation $\triangle B_{(p,q)}=0$, is provided by the single contribution of the subspace $\mathcal{H}_{p,q}$
\begin{equation}
\mathcal{Z}^B_{p,q}(\beta)=t_{p,q}(\beta)\;.
\end{equation}

We can now examine the structure of the effective action for the $\beta_{(p,d-p)}$ forms, that obey a null trace condition $\Tr\beta_{(p,d-p)}=0$ besides the Laplace-type equation. By using the same procedure of \eqref{B form exact} we get
\begin{equation}\label{beta form exact}
\begin{split}
\mathcal{Z}^\beta_{p,d-p}(\beta)&=-\frac12\oint_{\gamma}\frac{dz}{2\pi iz}\oint_{\gamma}\frac{dw}{2\pi iw}\frac{(z-w)^2}{zw}\,\Tr\Big[z^{(J_1^1-s)}w^{(J_2^2-s)}{\rm e}^{-\beta H}\Big]\\
&=-\frac12\oint_{\gamma}\frac{dz}{2\pi iz}\oint_{\gamma}\frac{dw}{2\pi iw}\frac{(z-w)^2}{zw}\,\Tr\Big[z^{(N-p)}w^{(-\bar N+d-p)}{\rm e}^{-\beta H}\Big]\\
&=-\frac12\oint_{\gamma}\frac{dz}{2\pi iz}\oint_{\gamma}\frac{dw}{2\pi iw}\frac{(z-w)^2}{zw}\,\sum_{n,m=0}^dz^{n-p}w^{d-p-m}\,\Tr_{\mathcal{H}_{n,m}}\Big[{\rm e}^{-\beta H}\Big]\\
&=t_{p,d-p}(\beta)-\frac12\,t_{p-1,d-p-1}(\beta)-\frac12\,t_{p+1,d-p+1}(\beta)\;.
\end{split}
\end{equation}
As it is natural, one has to subtract from the contribution of $(p,d-p)$-forms one term referring to the trace. The other one is effectively the same, since we will show in a while that $t_{p,q}=t_{d-q,d-p}$. However, its presence in the form $t_{p+1,d-p+1}$, that will be crucial in examining $\alpha$ forms, is due to the dual constraints we have in Dirac quantization: $J_2^1=\Tr$ and $J^2_1=g\wedge$, that are equivalent up to topological mismatches.
In the present theory no topological issues arise, indeed $t_{p+1,d-p+1}=t_{p-1,d-p-1}$, and one finally has
\begin{equation}
\mathcal{Z}^\beta_{p,d-p}(\beta)=t_{p,d-p}(\beta)-t_{p-1,d-p-1}(\beta)=\mathcal{Z}^B_{p,d-p}(\beta)-\mathcal{Z}^B_{p-1,d-p-1}(\beta)\;.
\end{equation}

We are now ready to deal with the more complicated models, \emph{i.e.} $A$ and $\alpha$, where supersymmetry is gauged. Here, as shown in \eqref{Z structure explained}, the additional poles
at $z=-1$ and $w=-1$ do arise, and one has to use the regulated contours discussed in the previous section.
Hence, by taking into account the residue at $z=0$ and both residues at $w=0$ and $w=-1$, one obtains for $A_{(p,q)}$ forms
\begin{equation}\label{A form exact}
\begin{split}
\mathcal{Z}^A_{p,q}(\beta)&=\oint_{\gamma^-}\frac{dz}{2\pi iz}\oint_{\gamma^+}\frac{dw}{2\pi iw}\frac{z}{(z+1)^2}\frac{w}{(w+1)^2}\,\Tr\Big[z^{(J_1^1-s_1)}w^{(J_2^2-s_2)}{\rm e}^{-\beta H}\Big]\\
&=\oint_{\gamma^-}\frac{dz}{2\pi iz}\oint_{\gamma^+}\frac{dw}{2\pi iw}\frac{z}{(z+1)^2}\frac{w}{(w+1)^2}\,\Tr\Big[z^{(N-p-1)}w^{(-\bar N+q+1)}{\rm e}^{-\beta H}\Big]\\
&=\oint_{\gamma^-}\frac{dz}{2\pi iz}\oint_{\gamma^+}\frac{dw}{2\pi iw}\frac{z}{(z+1)^2}\frac{w}{(w+1)^2}\,\sum_{n,m=0}^dz^{n-p-1}w^{q+1-m}\,\Tr_{\mathcal{H}_{n,m}}\Big[{\rm e}^{-\beta H}\Big]\\
&=\sum_{n=0}^p\sum_{m=0}^q(-)^{p-n}(-)^{q-m}(p+1-n)(q+1-m)t_{n,m}(\beta)\;,
\end{split}
\end{equation}
where we recall that $s_1=p+1-d/2$ and $s_2=d/2-q-1$. The structure of the effective action for $A_{(p,q)}$ forms, dictated by gauge symmetry, is then displayed in terms of the building blocks $t_{n,m}=\mathcal{Z}^B_{n,m}$
\begin{equation}
\mathcal{Z}^A_{p,q}(\beta)=\sum_{n=0}^p\sum_{m=0}^q(-)^{n+m}(n+1)(m+1)\mathcal{Z}^B_{p-n,q-m}(\beta)\;,
\end{equation}
where we changed variables $n\to p-n$ and $m\to q-m$ to cast it in this simple form.

The last model we are going to deal with concerns the $\alpha$ $(p,d-2-p)$-forms. Their field equations are analogous, apart from the trace term, to the equations for $A_{(p,q)}$ forms, and indeed the effective action
is recast neatly in terms of the $Z^A$
\begin{equation}\label{alpha form exact}
\begin{split}
&\mathcal{Z}^\alpha_{p,d-2-p}(\beta)=-\frac12\oint_{\gamma^-}\frac{dz}{2\pi iz}\oint_{\gamma^+}\frac{dw}{2\pi iw}\frac{z}{(z+1)^2}\frac{w}{(w+1)^2}\frac{(z-w)^2}{zw}\,\Tr\Big[z^{(J_1^1-s)}w^{(J_2^2-s)}{\rm e}^{-\beta H}\Big]\\
&=-\frac12\oint_{\gamma^-}\frac{dz}{2\pi iz}\oint_{\gamma^+}\frac{dw}{2\pi iw}\frac{z}{(z+1)^2}\frac{w}{(w+1)^2}\frac{(z-w)^2}{zw}\,\Tr\Big[z^{(N-p-1)}w^{(-\bar N+d-p-1)}{\rm e}^{-\beta H}\Big]\\
&=-\frac12\oint_{\gamma^-}\frac{dz}{2\pi iz}\oint_{\gamma^+}\frac{dw}{2\pi iw}\frac{z}{(z+1)^2}\frac{w}{(w+1)^2}\frac{(z-w)^2}{zw}\,\sum_{n,m=0}^dz^{n-p-1}w^{d-p-1-m}\,\Tr_{\mathcal{H}_{n,m}}\Big[{\rm e}^{-\beta H}\Big]\\
&=\sum_{n=0}^p\sum_{m=0}^{d-2-p}(-)^{p-n}(-)^{d-p-m}(p+1-n)(d-p-1-m)t_{n,m}(\beta)\\
&-\frac12\sum_{n=0}^{p-1}\sum_{m=0}^{d-3-p}(-)^{p-1-n}(-)^{d-p-3-m}(p-n)(d-p-2-m)t_{n,m}(\beta)\\
&-\frac12\sum_{n=0}^{p+1}\sum_{m=0}^{d-1-p}(-)^{p+1-n}(-)^{d-p-1-m}(p+1-n)(d-p-m)t_{n,m}(\beta)\;.
\end{split}
\end{equation}
By comparing with \eqref{A form exact}, we immediately recover from the last lines the effective actions of the $A_{(p,q)}$ forms, in terms of which the structure of $\mathcal{Z}^\alpha$ becomes apparent
\begin{equation}
\mathcal{Z}^\alpha_{p,d-2-p}(\beta)=\mathcal{Z}^A_{p,d-2-p}(\beta)-\frac12\,\mathcal{Z}^A_{p-1,d-3-p}(\beta)-\frac12\,\mathcal{Z}^A_{p+1,d-1-p}(\beta)\;.
\end{equation}
As in the case of $\beta$ forms, we see that the effective action is given by the corresponding contribution of the model without trace constraints \emph{i.e.} $\mathcal{Z}^A_{p,d-2-p}$, to which one has to subtract the terms corresponding to the trace and its dual. In this context, however, Hodge duality is affected by topological mismatches, so that $\mathcal{Z}^A_{p-1,d-3-p}\neq\mathcal{Z}^A_{p+1,d-1-p}$ and no further simplifications occur. This justifies the fact, already discussed in the previous section, that $\mathcal{Z}^\alpha_{0,d-2}$ differs from the naive expectation
of being identical to $\mathcal{Z}^A_{0,d-2}$: the difference is given by a topological quantity
\begin{equation}
\mathcal{Z}^\alpha_{0,d-2}(\beta)=\mathcal{Z}^A_{0,d-2}(\beta)-\frac12\,\mathcal{Z}^{A\,\text{top}}_{1,d-1}(\beta)\;.
\end{equation}

\subsection{Dualities and topological mismatches}

Having analyzed the structure of our effective actions in terms of the building blocks $t_{n,m}(\beta)\equiv\mathcal{Z}^B_{n,m}(\beta)$, we can now turn to study exact duality relations in the four models at hand. It will turn out that the symmetry under complex conjugation is always preserved, thanks to the choice of regulated contours $\gamma^-$ and $\gamma^+$. On the other hand, invariance under Hodge duality suffers a topological mismatch in the $A$ model, though it is preserved in the $B$ model. The remaining cases, $\alpha$ and $\beta$, do not show any mismatch, since Hodge duality acts trivially and does not even transform the fields.
 Hence, let us focus on the $A$ and $B$ forms.

We recall that the operators determining the form degree are given by
\begin{equation}
J^1_1=\frac12[\psi^\mu_1, \bar\psi_\mu^1]=N-\frac{d}{2}\;,\quad J^2_2=\frac12[\psi^\mu_2, \bar\psi_\mu^2]=-\bar N+\frac{d}{2}
\end{equation}
and that the Chern-Simons couplings differ for the  $A$ and the $B$ models
\begin{equation}\label{Chern-Simons couplings}
\begin{split}
s_1 &= p+1-\frac{d}{2}\;,\quad s_2=\frac{d}{2}-q-1\;,\quad A_{(p,q)}\,\text{forms}\\[2mm]
s_1 &= p-\frac{d}{2}\;,\qquad \ \  s_2=\frac{d}{2}-q\;,\qquad \ \ B_{(p,q)}\,\text{forms}\;.
\end{split}
\end{equation}
It is useful for the present analysis to recall also the gauge fixed fermionic action, that is common to the two models and reads
\begin{equation}\label{fermion action}
S^{\text{gf}}_{\text{fermion}}=\frac1\beta\int_0^1d\tau\Big[\bar\psi_\mu^iD_\tau\psi^\mu_i-\frac12 R_{\mu\bar\nu\lambda\bar\sigma}\,\psi^\mu\cdot\bar\psi^{\bar\nu}\psi^\lambda\cdot\bar\psi^{\bar\sigma}+i\phi\big(\bar\psi^1_\mu\psi_1^\mu+s_1\big)
+i\theta\big(\bar\psi^2_\mu\psi_2^\mu+s_2\big)\Big]\;.
\end{equation}
We start with the simpler $B_{(p,q)}$ forms, where one has $J^1_1-s_1=N-p$ and $J^2_2-s_2=-\bar N+q$. Under complex conjugation (c.c.) one has to swap holomorphic and anti-holomorphic indices, effectively exchanging $p\leftrightarrow q$. To realize this in the worldline theory it is necessary to exchange the Chern-Simons couplings: $s_1\stackrel{\text{c.c.}}{\longleftrightarrow}-s_2$, as it is evident from \eqref{Chern-Simons couplings}. By looking at the action \eqref{fermion action} it is clear that transforming the variables as $\psi^\mu_1\leftrightarrow\bar\psi_\mu^2$, $\bar\psi_\mu^1\leftrightarrow\psi^\mu_2$ and $\phi\leftrightarrow-\theta$, one brings the action back to its original form. This is perfectly legitimate, since it amounts to a change of dummy integration variables in the path integral. In the mixed formalism we are employing, the transformation $\psi^\mu_1\leftrightarrow\bar\psi_\mu^2$, $\bar\psi_\mu^1\leftrightarrow\psi^\mu_2$ produces $J_1^1\leftrightarrow-J^2_2$, and $\phi\leftrightarrow-\theta$ corresponds to $z\leftrightarrow\frac1w$. At the level of the effective action, the steps are as follows
\begin{equation}\label{B form cc}
\begin{split}
\mathcal{Z}^B_{q,p}(\beta)&=\oint_{\gamma}\frac{dz}{2\pi iz}\oint_{\gamma}\frac{dw}{2\pi iw}\,\Tr\Big[z^{(J_1^1+s_2)}w^{(J_2^2+s_1)}{\rm e}^{-\beta H}\Big]\\
&=\oint_{\gamma}\frac{dz}{2\pi iz}\oint_{\gamma}\frac{dw}{2\pi iw}\,\Tr\Big[z^{(-J_2^2+s_2)}w^{(-J_1^1+s_1)}{\rm e}^{-\beta H}\Big]\\
&=\oint_{\gamma}\frac{dw'}{2\pi iw'}\oint_{\gamma}\frac{dz'}{2\pi iz'}\Tr\Big[{w'}^{(J_2^2-s_2)}{z'}^{(J_1^1-s_1)}{\rm e}^{-\beta H}\Big]=\mathcal{Z}^B_{p,q}(\beta)\;.
\end{split}
\end{equation}
In the first line we wrote the effective action for the complex conjugated $(q,p)$-form, \emph{i.e.} with couplings $(-s_2, -s_1)$ instead of $(s_1,s_2)$; in the second step we renamed the fermionic variables, producing $J^1_1\leftrightarrow-J^2_2$, and in the last one we performed the change of variables $z=\frac{1}{w'}$, $w=\frac{1}{z'}$, that finally brought the effective action back to the original form,
thus proving the symmetry under complex conjugation.

To analyze the effects of Hodge duality on $B_{(p,q)}$ we proceed along similar lines. The Hodge transformation tells us to exchange $p\leftrightarrow d-q$, that in terms of Chern-Simons couplings is realized by $s_1\stackrel{\star}{\longleftrightarrow}s_2$. In order to undo the exchange and prove the symmetry one has to transform $\psi^\mu_1\leftrightarrow\psi^\mu_2$, $\bar\psi_\mu^1\leftrightarrow\bar\psi_\mu^2$ and $\phi\leftrightarrow\theta$, that is $z\leftrightarrow w$. This sends $J^1_1\leftrightarrow J^2_2$, and one can see the invariance of the effective action
\begin{equation}\label{B form hodge}
\begin{split}
\mathcal{Z}^B_{d-q,d-p}(\beta)&=\oint_{\gamma}\frac{dz}{2\pi iz}\oint_{\gamma}\frac{dw}{2\pi iw}\,\Tr\Big[z^{(J_1^1-s_2)}w^{(J_2^2-s_1)}{\rm e}^{-\beta H}\Big]\\
&=\oint_{\gamma}\frac{dz}{2\pi iz}\oint_{\gamma}\frac{dw}{2\pi iw}\,\Tr\Big[z^{(J_2^2-s_2)}w^{(J_1^1-s_1)}{\rm e}^{-\beta H}\Big]=\mathcal{Z}^B_{p,q}(\beta)\;.
\end{split}
\end{equation}
Again, we wrote first the Hodge dual effective action by using the coupling $(s_2,s_1)$ instead of $(s_1,s_2)$; then we performed the exchange between fermionic species, sending $J^1_1\leftrightarrow J^2_2$, and obtained immediately the original expression, due to the explicit symmetry between the $z$ and $w$ integrals.
This last symmetry will be missing when dealing with the $A$ model. This argument proves the invariance of the effective action under Hodge duality. Together with complex conjugation it ensures the exact validity of the following identities
\begin{equation}\label{t coefficients symmetries}
\begin{split}
\mathcal{Z}^B_{p,q}(\beta)&=\mathcal{Z}^B_{q,p}(\beta)=\mathcal{Z}^B_{d-q,d-p}(\beta)=\mathcal{Z}^B_{d-p,d-q}(\beta)\;,\quad\text{or}\\ t_{n,m}(\beta)&=t_{m,n}(\beta)=t_{d-m,d-n}(\beta)=t_{d-n,d-m}(\beta)\;.
\end{split}
\end{equation}

We are now going to investigate the more interesting case of $(p,q)$-form gauge fields $A_{(p,q)}$ where, due to the different Chern-Simons couplings, one has $J^1_1-s_1=N-p-1$ and $J^2_2-s_2=-\bar N+q+1$. We shall study first the action of complex conjugation. The transformed effective action is written by using the Chern-Simons couplings $(-s_2, -s_1)$, and the change of variables to undo the transformation is the same
as before
\begin{equation}\label{A form cc}
\begin{split}
\mathcal{Z}^A_{q,p}(\beta)&=\oint_{\gamma^-}\frac{dz}{2\pi iz}\oint_{\gamma^+}\frac{dw}{2\pi iw}\frac{z}{(z+1)^2}\frac{w}{(w+1)^2}\,\Tr\Big[z^{(J_1^1+s_2)}w^{(J_2^2+s_1)}{\rm e}^{-\beta H}\Big]\\
&=\oint_{\gamma^-}\frac{dz}{2\pi iz}\oint_{\gamma^+}\frac{dw}{2\pi iw}\frac{z}{(z+1)^2}\frac{w}{(w+1)^2}\,\Tr\Big[z^{(-J_2^2+s_2)}w^{(-J_1^1+s_1)}{\rm e}^{-\beta H}\Big]\\
&=\oint_{\gamma^+}\frac{dw'}{2\pi iw'}\oint_{\gamma^-}\frac{dz'}{2\pi iz'}\frac{w'}{(w'+1)^2}\frac{z'}{(z'+1)^2}\,\Tr\Big[{w'}^{(J_2^2-s_2)}{z'}^{(J_1^1-s_1)}{\rm e}^{-\beta H}\Big]=\mathcal{Z}^A_{p,q}(\beta)\;.
\end{split}
\end{equation}
The important difference from the $B$ form computation is the presence of regulated contours $\gamma^-$ and $\gamma^+$: when performing the change of variables $z=\frac{1}{w'}$, $w=\frac{1}{z'}$, each regulated contour maps to the other one $\gamma^-\leftrightarrow\gamma^+$. In this circumstance the new integrals coincide with the original form of the effective action, and no mismatch appears, proving invariance under complex conjugation
\begin{equation}
\mathcal{Z}^A_{p,q}(\beta)=\mathcal{Z}^A_{q,p}(\beta)\;.
\end{equation}
Alternatively, this could have been proved using the manifest symmetry under $p\leftrightarrow q$ of \eqref{A form exact} and \eqref{t coefficients symmetries}.

The action of Hodge duality is considerably more involved: the dual effective action is written with couplings $(s_2, s_1)$, but the regulated contours will be in the wrong order and a mismatch appears. Explicitly one has
\begin{equation}\label{A form hodge}
\begin{split}
\mathcal{Z}^A_{d-2-q,d-2-p}(\beta)&=\oint_{\gamma^-}\frac{dz}{2\pi iz}\oint_{\gamma^+}\frac{dw}{2\pi iw}\frac{z}{(z+1)^2}\frac{w}{(w+1)^2}\,\Tr\Big[z^{(J_1^1-s_2)}w^{(J_2^2-s_1)}{\rm e}^{-\beta H}\Big]\\
&=\oint_{\gamma^-}\frac{dz}{2\pi iz}\oint_{\gamma^+}\frac{dw}{2\pi iw}\frac{z}{(z+1)^2}\frac{w}{(w+1)^2}\,\Tr\Big[z^{(J_2^2-s_2)}w^{(J_1^1-s_1)}{\rm e}^{-\beta H}\Big]\\
&=\oint_{\gamma^+}\frac{dz'}{2\pi iz'}\oint_{\gamma^-}\frac{dw'}{2\pi iw'}\frac{z'}{(z'+1)^2}\frac{w'}{(w'+1)^2}\,\Tr\Big[z'^{(J_1^1-s_1)}w'^{(J_2^2-s_2)}{\rm e}^{-\beta H}\Big]
\end{split}
\end{equation}
where in the last line we simply renamed $z=w'$ and $w=z'$ for an immediate legibility.
Hence we see that the new effective action is identical to the original one in \eqref{A form exact} apart from the contour choice, that is inverted. Let us denote with $\gamma^0$ a small circle surrounding $-1$ in the complex $z$ or $w$ plane, so that $\gamma^+=\gamma^-+\gamma^0$. The mismatch between the two effective actions is then given by
\begin{equation}\label{A difference}
\begin{split}
\mathcal{Z}^A_{d-2-q,d-2-p}(\beta)&=\oint_{\gamma^+}\frac{dz}{2\pi iz}\oint_{\gamma^-}\frac{dw}{2\pi iw}\frac{z}{(z+1)^2}\frac{w}{(w+1)^2}\,\Tr\Big[z^{(J_1^1-s_1)}w^{(J_2^2-s_2)}{\rm e}^{-\beta H}\Big]\\[2mm]
&=\mathcal{Z}^A_{p,q}(\beta)+\left[\oint_{\gamma^0}\oint_{\gamma^+}-\oint_{\gamma^-}\oint_{\gamma^0}-\oint_{\gamma^0}\oint_{\gamma^0}\right]\frac{dz}{2\pi iz}\frac{dw}{2\pi iw}\\
&\hspace{15mm}\times\frac{z}{(z+1)^2}\frac{w}{(w+1)^2}\,\Tr\Big[z^{(J_1^1-s_1)}w^{(J_2^2-s_2)}{\rm e}^{-\beta H}\Big]\\[2mm]
&=\mathcal{Z}^A_{p,q}(\beta)+\left[\sum_{n,m=0}^d-\sum_{n=0}^p\sum_{m=0}^d-\sum_{n=0}^d\sum_{m=0}^q\right]f_{n,m}(p,q;\beta)\;,
\end{split}
\end{equation}
where we defined
\begin{equation}
f_{n,m}(p,q;\beta)\equiv(-)^{n-p}(-)^{m-q}(p+1-n)(q+1-m)t_{n,m}(\beta)\;.
\end{equation}
To obtain the last line we computed the residues as dictated by the contours, recalling that $\Tr\Big[z^{(J_1^1-s_1)}w^{(J_2^2-s_2)}{\rm e}^{-\beta H}\Big]=\sum_{n,m=0}^dz^{n-p-1}w^{-m+q+1}t_{n,m}$. We shall demonstrate in Appendix \ref{app:topology} that the difference $\mathcal{Z}^A_{d-2-q,d-2-p}(\beta)-\mathcal{Z}^A_{p,q}(\beta)$ is purely topological, indeed given by linear combinations of quantum mechanical indices of the form $\Tr(-1)^F$ and $\Tr(-1)^FF$, where $F$ is a suitable fermion number operator. The main identifications we need, to be proven in
Appendix \ref{app:topology}, are
\begin{equation}\label{identifications}
\begin{split}
&\sum_{m=0}^d(-)^mt_{n,m}(\beta) = \Tr_{\mathcal{H}_n}(-1)^{\bar N}= \text{ind}\left(\Omega^{n,0},\bar\de\right)\;,\quad\text{and complex conjugate}\\
&\sum_{n,m=0}^d(-)^{n+m}t_{n,m}(\beta) = \Tr_{\mathcal{H}}(-1)^{N+\bar N}= \chi(\mathcal{M})\;,
\end{split}
\end{equation}
where $\mathcal{H}_n=\bigoplus_{m=0}^d\mathcal{H}_{n,m}$ is the Hilbert subspace with fixed $N=n$, $\chi(\mathcal{M})$ is the Euler characteristics of our manifold, and $\text{ind}\left(\Omega^{n,0},\bar\de\right)$ denotes the Dolbeault index twisted by the holomorphic vector bundle of $(n,0)$-forms. By putting together the various contributions from \eqref{A difference}, and using \eqref{identifications}, we finally get the relation between the effective actions of Hodge dual forms
\begin{equation}\label{mismatch}
\begin{split}
\mathcal{Z}^A_{d-2-q, d-2-p}(\beta)-\mathcal{Z}^A_{p, q}(\beta)&=(-)^{q+d}\mathcal{Z}^{A\,\text{top}}_{p, d-1}(\beta)+(-)^{p+d}\mathcal{Z}^{A\,\text{top}}_{d-1,q}(\beta)+(-)^{p+q}\mathcal{Z}^{A\,\text{top}}_{d-1, d-1}(\beta)\\
&+(d-1-p)(-)^{p+q}\sum_{m=0}^q(-)^m(q+1-m)\text{ind}\left(\Omega^{m,0},\bar\de\right)\\
&+(d-1-q)(-)^{p+q}\sum_{n=0}^p(-)^n(p+1-n)\text{ind}\left(\Omega^{n,0},\bar\de\right)\\
&+(-)^{p+q}\left[\Big(p+1-\frac{d}{2}\Big)\Big(q+1-\frac{d}{2}\Big)-\frac{d^2}{4}\right]\chi({\mathcal{M}})
\end{split}
\end{equation}
where the contributions of non-propagating top forms (\emph{i.e.} forms whith $p$ or $q$ exceeding $d-2$) have been read down from the explicit formula \eqref{A form exact} for $p=d-1$ and/or $q=d-1$. It may appear rather obscure that such top forms, although obviously not propagating, are in fact topological. To render it manifest, we stress that they have a genuine topological interpretation, since they coincide with analytic Ray-Singer torsions of the $\bar\de$-operator \cite{Ray:1973sb}, as it will be shown in Appendix \ref{app:topology}, namely
\begin{equation}
\begin{split}
&\int_0^\infty\frac{d\beta}{\beta}\sum_{m=0}^d(-)^m\,m\,t_{n,m}(\beta)=\int_0^\infty\frac{d\beta}{\beta}\Tr_{\mathcal{H}_n}\Big[(-1)^{\bar N}\,\bar N\,{\rm e}^{-\beta H}\Big]=2\ln T_n(\mathcal{M})\quad\Rightarrow\\
&Z^{A\,\text{top}}_{p,d-1}=2\sum_{n=0}^p(-)^{n+1}(n+1)\ln T_{d-p+n}(\mathcal{M})\;,
\end{split}
\end{equation}
where $T_n(\mathcal{M})$ denotes the $\bar\de$-torsion obtained by summing in $m$ the contributions of $(n,m)$-forms at fixed $n$. We stress here that, unlike the case of Dolbeault indices, the identification with the Ray-Singer torsions is valid at the level of effective actions $Z$ and not of densities $\mathcal{Z}$.
A similar phenomenon is present for ordinary differential forms \cite{Schwarz:1984wk}.

These topological mismatches, that arise as obstructions in defining uniquely the Hodge dual gauge fields on the whole manifold, are visible in our computations only in $d=3$ and $d=2$, since in higher dimensions they appear in higher order Seeley-DeWitt coefficients. Nevertheless, also in these particular cases we can address very non-trivial checks.
For instance, in $d=3$ we can check \eqref{mismatch} for $p=q=0$ and for $p=0$, $q=1$:
\begin{equation}\label{check d=3}
\begin{split}
&d=3\\[2mm]
&\mathcal{Z}^A_{1,1}(\beta)-\mathcal{Z}^A_{0,0}(\beta)=\mathcal{Z}^{A\,\text{top}}_{2,2}(\beta)-2\mathcal{Z}^{A\,\text{top}}_{2,0}(\beta)\\[2mm]
&\mathcal{Z}^{A\,\text{top}}_{2,0}(\beta)-\mathcal{Z}^{A\,\text{top}}_{2,2}(\beta)-\mathcal{Z}^{A\,\text{top}}_{2,1}(\beta)=0\;,
\end{split}
\end{equation}
where we have omitted the index contributions since they are of order $\beta^3$. From \eqref{A form coeff d=3} we can read off the SDW coefficients of $A_{(0,0)}$ and $A_{(1,1)}$. Their difference yields
\begin{equation}
\begin{split}
&d=3\\[2mm]
&A_{(1,1)}-A_{(0,0)}\to\left(0;\,0;\,\frac{5}{12};\,-\frac{2}{3};\,\frac{1}{4};\,0\right)\;,
\end{split}
\end{equation}
while from \eqref{Z A} we find the contributions of the top forms as
\begin{equation}
\begin{split}
&d=3\\[2mm]
&A_{(2,2)}\to\left(0;\,0;\,\frac{1}{2};\,-1;\,\frac{1}{2};\,0\right)\;,\\[2mm]
&A_{(2,0)}\to\left(0;\,0;\,\frac{1}{24};\,-\frac{1}{6};\,\frac{1}{8};\,0\right)\;,\\[2mm]
&A_{(2,1)}\to\left(0;\,0;\,-\frac{11}{24};\,\frac{5}{6};\,-\frac{3}{8};\,0\right)\;,
\end{split}
\end{equation}
and it is straightforward to see that they satisfy both relations in \eqref{check d=3}.

In two complex dimensions we can check \eqref{mismatch} at $p=q=0$, for which we get the following relation between topological quantities
\begin{equation}\label{check d=2}
\begin{split}
&d=2\\[2mm]
&\mathcal{Z}^{A\,\text{top}}_{1,1}(\beta)+2\mathcal{Z}^{A\,\text{top}}_{1,0}(\beta)+2\,\text{ind}\left(\bar\de\right)-\chi(\mathcal{M})=0\;.
\end{split}
\end{equation}
Again, the top form contributions can be computed from \eqref{Z A}:
\begin{equation}
\begin{split}
&d=2\\[2mm]
&A_{(1,1)}\to\left(0;\,1;\,\frac{1}{2};\,-\frac{11}{12};\,\frac{5}{12};\,\frac{1}{12}\right)\;,\\[2mm]
&A_{(1,0)}\to\left(0;\,-\frac12;\,-\frac{1}{24};\,\frac{1}{8};\,-\frac{1}{12};\,-\frac{1}{24}\right)\;,
\end{split}
\end{equation}
and they give, as sum of the effective action densities,
\begin{equation}\label{piece one}
\mathcal{Z}^{A\,\text{top}}_{1,1}(\beta)+2\mathcal{Z}^{A\,\text{top}}_{1,0}(\beta)=\frac{1}{4\pi^2}\int d^2x_0d^2\bar x_0 g(x_0)\,\Big[\frac{5}{12}\,
R_{\mu\bar\nu\lambda\bar\sigma}R^{\mu\bar\nu\lambda\bar\sigma}
-\frac23\,R_{\mu\bar\nu}R^{\mu\bar\nu}+\frac14\,R^2\Big].
\end{equation}
The index of the Dolbeault operator is given by \eqref{index d=2} and in components reads
\begin{equation}\label{piece two}
\begin{split}
\text{ind}\left(\bar\de\right)&=\frac{1}{32\pi^2}\int_{\mathcal{M}}\Big[\frac13\,\text{tr}\,\mathcal{R}\wedge\mathcal{R}-\text{tr}\,\mathcal{R}\wedge\text{tr}\,\mathcal{R}\Big]\\
&=\frac{1}{32\pi^2}\int d^2x_0d^2\bar x_0 g(x_0)\,\Big[\frac{1}{3}\,
R_{\mu\bar\nu\lambda\bar\sigma}R^{\mu\bar\nu\lambda\bar\sigma}
-\frac43\,R_{\mu\bar\nu}R^{\mu\bar\nu}+R^2\Big],
\end{split}
\end{equation}
with the K\"ahler curvature two-form being $\mathcal{R}^\lambda{}_\sigma=R_{\mu\bar\nu}{}^\lambda{}_\sigma dx^\mu\wedge d\bar x^{\bar\nu}$.
The Euler characteristics is given by
\begin{equation}\label{piece three}
\begin{split}
\chi(\mathcal{M})&=\frac{1}{32\pi^2}\int_{\mathcal{M}}\Big[\epsilon^{ABCD}\mathcal{R}^{(G)}_{AB}\wedge\mathcal{R}^{(G)}_{CD}\Big]\\
&=\frac{1}{8\pi^2}\int d^2x_0d^2\bar x_0 g(x_0)\,\Big[R_{\mu\bar\nu\lambda\bar\sigma}R^{\mu\bar\nu\lambda\bar\sigma}
-2\,R_{\mu\bar\nu}R^{\mu\bar\nu}+R^2\Big],
\end{split}
\end{equation}
where $\mathcal{R}^{(G)}_{AB}$ denotes the curvature two-form built from the riemannian metric $G_{MN}$, and $A,B,..$ riemannian $SO(2d)$ flat indices.
Putting together \eqref{piece one}, \eqref{piece two} and  \eqref{piece three}, one easily verifies that the relation \eqref{check d=2} is indeed satisfied.
\section{Conclusions}

We have described several quantum theories of massless $(p,q)$-forms, with and without gauge symmetries.
We have employed a worldline approach that makes use of a U(2) spinning particle,
which is identified by a supersymmetric nonlinear sigma model with four supercharges suitably gauged.
In particular, we have studied effective actions of massless $(p,q)$-forms
on arbitrary K\"ahler spaces,  and studied duality relations.
For the  A-type models they include topological mismatches,
related to index densities and analytic torsions. Calculation of several heat kernel coefficients has been presented as well.
The physical motivations for studying such models are indirect, since a honest spacetime interpretation is prevented by the complex nature
of the target space (also choosing an appropriate signature, time must necessarily be complex to preserve the complex structure of the target manifold).
Nevertheless complex manifolds find useful applications in the context of string and/or  supersymmetric theories.
For sure, they offer a useful arena to test methods of quantum field theory,
including worldline approaches to theories on curved backgrounds and ideas form the current studies of higher spin fields.

\acknowledgments{We wish to thank Roberto Zucchini for valuable discussions. The work of FB was supported in part by the MIUR-PRIN contract 2009-KHZKRX. C.I. wishes to thank the Erwin Schr\"odinger Institute in Vienna for kind hospitality during the final stage of this work, and gratefully acknowledges the ``Universit\`a degli Studi G. Marconi'' in Rome for partial support.}
\vskip2cm

\vfill\eject

\appendix
\section{Notations and conventions}
\label{app:notations}

We list here the conventions and formulas of K\"ahler geometry we make use of in the main text.
When needed, we rewrite some geometric quantities by using both holomorphic coordinates $(x^\mu, \bar x^{\bar\mu})$, $\mu,\bar\mu=1,...,d$, and riemannian coordinates $X^M$, with $M=1,...,2d$.
We deal with K\"ahler manifolds $\mathcal{M}$ with $d$ complex dimensions, whose metric is specified by
\begin{equation}
ds^2 = G_{MN} dX^M dX^N = 2 g_{\mu\bar\nu} dx^\mu d\bar x^{\bar\nu}\;,
\end{equation}
while the integration measure reads
\begin{equation}
d\mu = \sqrt{\det{G_{MN}}}\, d^D X =  \det{g_{\mu\bar\nu}}\, d^d x  d^d\bar x
\end{equation}
with the convention
\begin{equation} \label{def-measure}
  d^d x  d^d\bar x \equiv i^d \prod_{\mu=1}^d dx^\mu \wedge d\bar x^{\bar \mu} \;.
\end{equation}
We also use the notation $ g\equiv\det{g_{\mu\bar\nu}}$ for the determinant of the K\"ahler metric.
On flat manifolds one can employ cartesian coordinates for which $G_{MN}=\delta_{MN}$ and
$g_{\mu\bar\nu}=\delta_{\mu\bar\nu}$.
We  relate real and complex coordinates by
\begin{equation}
x^\mu = \frac{1}{\sqrt{2}} (X^{2\mu-1} + i X^{2\mu}) \;, \quad \bar x^{\bar \mu} = \frac{1}{\sqrt{2}} (X^{2\mu-1} - i X^{2\mu}) \;,
\quad \mu=1,...,d\;.
\end{equation}

We come now to connections and curvatures. In holomorphic coordinates the nonzero Christoffel symbols are given, in terms of the metric, by
\begin{equation}
\Gamma^\mu_{\nu\lambda}=g^{\mu\bar\mu}\de_\nu g_{\lambda\bar\mu}\;,
\quad \Gamma^{\bar\mu}_{\bar\nu\bar\lambda}=g^{\mu\bar\mu}\de_{\bar\nu} g_{\bar\lambda\mu}\;,
\end{equation}
where $\de_\mu\equiv\frac{\de}{\de x^\mu}$ and $\de_{\bar\mu}\equiv\frac{\de}{\de\bar x^{\bar\mu}}$.
The nonzero components of the Riemann tensor read
\begin{equation}
R^\mu{}_{\nu\bar\sigma\lambda}=\de_{\bar\sigma}\Gamma^\mu_{\nu\lambda}\;,
\quad R^{\bar\mu}{}_{\bar\nu\sigma\bar\lambda}=\de_{\sigma}\Gamma^{\bar\mu}_{\bar\nu\bar\lambda}\;,
\end{equation}
while we define the Ricci tensor and the curvature scalar as
\begin{equation}
\begin{split}
R_{\mu\bar\nu} &= -R^\lambda{}_{\lambda\bar\nu\mu}=-\de_\mu\bar\Gamma_{\bar\nu}=-\de_{\bar\nu}\Gamma_\mu
=-\de_\mu\de_{\bar\nu}\ln g\;,\\[2mm]
R &= g^{\mu\bar\nu}R_{\mu\bar\nu}\;.
\end{split}
\end{equation}
The curvature two-forms for K\"ahler and riemannian connections is denoted by
\begin{equation}\label{curvature forms}
\mathcal{R}^\mu{}_\nu=R_{\lambda\bar\sigma}{}^\mu{}_\nu\,dx^\lambda\wedge d\bar x^{\bar\sigma}\;,\quad \mathcal{R}^{(G)}_{AB}=\frac12\,R_{MNAB}\,dX^M\wedge dX^N\;,
\end{equation}
and the volume forms for K\"ahler and riemannian metrics read
\begin{equation}
g(x,\bar x)\epsilon_{\mu_1...\mu_d}\epsilon_{\bar\nu_1...\bar\nu_d}\;,\quad \sqrt{G(X)}\epsilon_{M_1...M_{2d}}\;.
\end{equation}

The $(p,q)$-forms are written in components as
\begin{equation}
A_{(p,q)}=A_{\mu_1...\mu_p\bar\nu_1...\bar\nu_q}(x, \bar x)\,dx^{\mu_1}\wedge...\wedge dx^{\mu_p}\wedge d\bar x^{\bar\nu_1}...\wedge d\bar x^{\bar\nu_q}\;,
\end{equation}
but for convenience of notation, from now on, the wedge product is understood whenever a $dx^\mu$ or $d\bar x^{\bar\mu}$ is acting.
Number operators, trace operator and multiplication by the K\"ahler form act as
\begin{equation}
\begin{split}
N&=dx^\mu\frac{\de}{\de(dx^\mu)}\;,\quad \bar N=d \bar x^{\bar\nu}\frac{\de}{\de(d\bar x^{\bar\nu})}\\
\Tr&= g^{\mu\bar\nu}\frac{\de^2}{\de(dx^\mu)\de(d\bar x^{\bar\nu})}\;,\quad g\wedge = g_{\mu\bar\nu}dx^\mu d\bar x^{\bar\nu}\;.
\end{split}
\end{equation}
In defining Dolbeault operators and their adjoint we choose to denote by $\de^\dagger$ and $\bar\de^\dagger$ the actual adjoints of $\de$ and $\bar\de$, so that a minus sign appears in the divergences
\begin{equation}
\begin{split}
\de&=dx^\mu\de_\mu\;,\quad \bar\de=d\bar x^{\bar\mu}\de_{\bar\mu}\\
\de^\dagger&=-g^{\mu\bar\nu}\frac{\de}{\de(dx^\mu)}\nabla_{\bar\nu}\;,\quad \bar\de^\dagger=-g^{\mu\bar\nu}\frac{\de}{\de(d\bar x^{\bar\nu})}\nabla_{\mu}\;.
\end{split}
\end{equation}
The $U(d)$ ``Lorentz'' generators are given by
\begin{equation}\label{Ud}
M^{\mu\bar\nu}=g^{\nu\bar\nu}\,dx^\mu\frac{\de}{\de(dx^\nu)}-g^{\mu\bar\mu}\,d\bar x^{\bar\nu}\frac{\de}{\de(d\bar x^{\bar\mu})}\;.
\end{equation}
We denote by $\nabla^2$ the curved space laplacian
\begin{equation}
\nabla^2=G^{MN}\nabla_M\nabla_N=g^{\mu\bar\nu}\left(\nabla_\mu\nabla_{\bar\nu}+\nabla_{\bar\nu}\nabla_\mu\right)\;,
\end{equation}
while we reserve the symbol $\triangle$ for the full Laplace-Beltrami operator with Dolbeault normalization
\begin{equation}
\triangle=\frac{\nabla^2}{2}+\frac12\,R_{\mu\bar\nu\lambda\bar\sigma}\,M^{\mu\bar\nu}M^{\lambda\bar\sigma}=-\left(\de\de^\dagger+\de^\dagger\de\right)
=-\left(\bar\de\bar\de^\dagger+\bar\de^\dagger\bar\de\right)=-\frac12\left(dd^\dagger+d^\dagger d\right)
\end{equation}
so that the full quantum hamiltonian acts as $H=-\triangle$.

Finally, in order to compare the heat kernel coefficients computed in the present paper with those present in the literature,
we relate the quadratic terms in curvatures with riemannian normalization to those
with K\"ahler normalization
\begin{equation}
R_{(G)}\equiv g^{MN}R_{MN}=2R\;,\quad R_{MN}R^{MN}=2R_{\mu\bar\nu}R^{\mu\bar\nu}\;,
\quad R_{MNRS}R^{MNRS}=4R_{\mu\bar\nu\rho\bar\sigma}R^{\mu\bar\nu\rho\bar\sigma}\;.
\end{equation}
Note in particular that the curvature scalar with K\"ahler normalization is one half of the standard riemannian one.

\section{Propagators in TS regularization}\label{app:propagators}

We list here all the propagators in the continuum limit, along with their derivatives and equal time expressions, as well as their regulated expressions valid in Time Slicing regularization. All the TS prescriptions follow by considering the Dirac distributions acting as Kronecker deltas, and taking into account the regulated value $\theta(0)=\frac12$.

With $\t,\s\in[0,1]$ and denoting $\phi_i=(\phi,\theta)$ the propagators read
\begin{equation}
\begin{split}
&\Delta(\tau,\sigma) = \sigma(\tau-1)\,\theta(\tau-\sigma)+\tau(\sigma-1)\,\theta(\sigma-\tau)\;,\\
&\Delta_{\text{gh}}(\t,\s)= \delta(\t,\s)\;,\\
&\Delta_{\text{f}}(\tau-\sigma,\phi_i) = \frac{{\rm e}^{-i\phi_i (\tau-\sigma)}}{2\cos\frac{\phi_i}{2}}\big[{\rm e}^{i\frac{\phi_i}{2}}\theta(\tau-\sigma)-{\rm e}^{-i\frac{\phi_i}{2}}\theta(\sigma-\tau)\big]\;.
\end{split}
\end{equation}
Derivatives of the bosonic propagator in the continuum limit are
\begin{equation}
\begin{split}
\puntos{\Delta}(\t,\s)&=\s-\theta(\s-\t)\;,\quad\puntod{\Delta}(\t,\s)=\t-\theta(\t-\s)\;,\\
\puntods{\Delta}(\t,\s)&=1-\delta(\t,\s)\;,
\end{split}
\end{equation}
where left (right) dots indicate derivatives with respect to the first (second) variable.

In Time Slicing regularization the following relations hold
\begin{equation}
\begin{split}
&\puntods{\Delta}(\t,\s)+\Delta_{\text{gh}}(\t,\s)=1\;,\\
&\Delta_{\text{f}}(\tau-\sigma,\phi_i)\Delta_{\text{f}}(\s-\t,\phi_i)=-\frac14\,\cos^{-2}\frac{\phi_i}{2}\;,
\end{split}
\end{equation}
the first one ensuring that no products of delta distributions will ever arise in perturbative calculations.
Finally, equal time expressions of the propagators and of their derivatives are obtained from the aforementioned TS prescriptions, and given explicitly by
\begin{equation}
\begin{split}
\Delta(\t,\t)&=\t(\t-1)\;,\quad \puntos{\Delta}(\t,\t)=\puntod{\Delta}(\t,\t)=\t-\frac12\;,\\
\Delta_{\text{f}}(0,\phi_i)&=\frac{i}{2}\,\tan\frac{\phi_i}{2}\;.
\end{split}
\end{equation}

\section{Topological quantities}\label{app:topology}

We dedicate this appendix to demonstrate the topological identifications we made use of in the main text.
Let us start with the Dolbeault indices. We mentioned that our Hilbert space can be viewed as a direct sum of subspaces with fixed fermion numbers
\begin{equation}
\mathcal{H}=\bigoplus_{n,m=0}^d\mathcal{H}_{n,m}\;.
\end{equation}
Since $N$, $\bar N$ and $H$ are self-adjoint and commute with each other, they can be diagonalized simultaneously to provide
a basis of eigenstates for each subspace $\mathcal{H}_{n,m}$. The same is true if one sums over one fermion number, obtaining the Hilbert subspace with only one fermion number fixed
\begin{equation}
\mathcal{H}_{n}=\bigoplus_{m=0}^d\mathcal{H}_{n,m}\;.
\end{equation}
We can focus now on one alternate sum of $t_{n,m}$ coefficients, that can be rewritten as
\begin{equation}
\sum_{m=0}^d(-)^mt_{n,m}(\beta)=\sum_{m=0}^d(-)^m \Tr_{\mathcal{H}_{n,m}}\Big[{\rm e}^{-\beta H}\Big]=\Tr_{\mathcal{H}_{n}}\Big[(-)^{\bar N}{\rm e}^{-\beta H}\Big]\;.
\end{equation}
We notice that the part of the superalgebra generated by $Q_2$, $\bar Q^2$ and $H$ acts separately within each $\mathcal{H}_n$. The Witten index argument is then applicable, and tells us that for each $H$-eigenstate $\ket{j}$ with energy $\epsilon_j$, there is another eigenstate $Q_2\ket{j}$ with the same energy and opposite value of $(-)^{\bar N}$, and hence only zero modes of the hamiltonian survive in the trace written above. Since our hamiltonian coincides with the Dolbeault laplacian, the trace effectively counts with alternate signs the numbers of harmonic $(n,m)$-forms, \emph{i.e.} the Hodge numbers $h^{n,m}$
\begin{equation}\label{t sum as h sum}
\begin{split}
\sum_{m=0}^d(-)^mt_{n,m}(\beta)&=\Tr_{\mathcal{H}_{n}}\Big[(-)^{\bar N}{\rm e}^{-\beta H}\Big]=\sum_{m=0}^d(-)^m\left(\text{number of harmonic (n,m)-forms}\right)\\
&=\sum_{m=0}^d(-)^mh^{n,m}(\mathcal{M})\;,
\end{split}
\end{equation}
that for $n=0$ would be the Dolbeault index: $\sum_{m=0}^d(-)^mh^{0,m}=\text{ind}(\bar\de)$. The generalization we need requires to consider Dolbeault operators acting on a holomorphic vector bundle $V$. The corresponding Hodge numbers provide the index of the twisted Dolbeault operator \cite{Nakahara}
\begin{equation}\label{twisted dolbeault}
\sum_{m=0}^d(-)^mh^{0,m}(V,\mathcal{M})=\text{ind}\left(V,\bar\de\right)\;.
\end{equation}
For our purposes, the role of $V$ is played by the bundle of $(n,0)$-forms: $V=\Omega^{n,0}$, and the Hodge numbers can be identified as $h^{n,m}(\mathcal{M})=h^{0,m}(\Omega^{n,0},\mathcal{M})$,
providing us with the Dolbeault index twisted by $(n,0)$-forms
\begin{equation}\label{t as index}
\sum_{m=0}^d(-)^mt_{n,m}(\beta)=\sum_{m=0}^d(-)^mh^{n,m}(\mathcal{M})=\text{ind}\left(\Omega^{n,0},\bar\de\right)\;.
\end{equation}
By symmetry under complex conjugation the same result is valid exchanging the roles of $n$ and $m$,
and substituting $\bar \de$ by $\de$.
The explicit formula for it is given as usual by the Atiyah-Singer index theorem, that in the present case reduces to the Hirzebruch-Riemann-Roch theorem
\begin{equation}\label{HirzRiemRoch}
\text{ind}\left(V,\bar\de\right)=\int_{\mathcal{M}}\text{Td}(T\mathcal{M}^+)\text{ch}(V)\;,
\end{equation}
where $\text{Td}(T\mathcal{M}^+)$ is the Todd class of the holomorphic tangent space bundle
\begin{equation}
\text{Td}(T\mathcal{M}^+)=\det\left(\frac{i\mathcal{R}/2\pi}{1-{\rm e}^{-i\mathcal{R}/2\pi}}\right)\;,
\end{equation}
with $\mathcal{R}$ the K\"ahler curvature form \eqref{curvature forms}, and
\begin{equation}
\text{ch}(V)=\text{tr}\,{\rm e}^{i\mathcal{F}_V/2\pi}
\end{equation}
 the Chern character of the vector bundle $V$, $\mathcal{F}_V$ being its curvature two-form. Since our $V$ consists of $(n,0)$-forms, the formula becomes
\begin{equation}
\text{ind}\left(\Omega^{n,0},\bar\de\right)=\int_{\mathcal{M}}\text{Td}(T\mathcal{M}^+)\text{ch}\left(\bigwedge^nT^*\mathcal{M}\right)\;.
\end{equation}
The Chern character for the anti-symmetric product can be computed by means of curvature eigenvalues $x_i$ (see \cite{Nakahara}) as
\begin{equation}
\text{ch}\left(T\mathcal{M}\right)=\sum_i{\rm e}^{x_i}\rightarrow \text{ch}\left(\bigwedge^nT^*\mathcal{M}\right)=\sum_{i_1<i_2...<i_n}{\rm e}^{-x_{i_1}}{\rm e}^{-x_{i_2}}...{\rm e}^{-x_{i_n}}\;,
\end{equation}
and can be reexpressed as
\begin{equation}\label{chern character}
\begin{split}
\text{ch}\left(\bigwedge^nT^*\mathcal{M}\right)&=\frac{1}{n!}\left\{\left(\text{tr}\,{\rm e}^{-i\mathcal{R}/2\pi}\right)^n-\binom{n}{2}\left(\text{tr}\,{\rm e}^{-i\mathcal{R}/2\pi}\right)^{n-2}\left(\text{tr}\,{\rm e}^{-2i\mathcal{R}/2\pi}\right)\right.\\
&\left.+\sum_{k=3}^n\binom{n}{k}\frac{2k-n-1}{n-k+1}\left(\text{tr}\,{\rm e}^{-i\mathcal{R}/2\pi}\right)^{n-k}\left(\text{tr}\,{\rm e}^{-ki\mathcal{R}/2\pi}\right)\right\}\;.
\end{split}
\end{equation}
For a check we are able to perform in section \ref{sec dualities} we need $\text{ind}(\bar\de)$ in two complex dimensions
\begin{equation}
\text{ind}\left(\bar\de\right)=\int_{\mathcal{M}}\text{Td}(T\mathcal{M}^+)\;.
\end{equation}
The Todd class can be expanded as
\begin{equation}
\text{Td}(T\mathcal{M}^+)=1+\frac12\,\text{tr}\,\left(\frac{i\mathcal{R}}{2\pi}\right)-\frac{1}{24}\,\text{tr}\,\left(\frac{i\mathcal{R}}{2\pi}\right)^2+\frac18\,
\left[\text{tr}\,\left(\frac{i\mathcal{R}}{2\pi}\right)\right]^2+\mathcal{O}(\mathcal{R}^3)\;,
\end{equation}
and the index in $d=2$ is
\begin{equation}\label{index d=2}
\begin{split}
&\text{ind}\left(\bar\de\right)=\frac{1}{32\pi^2}\int_{\mathcal{M}}\Big[\frac13\,\text{tr}\,\mathcal{R}\wedge\mathcal{R}-\text{tr}\,\mathcal{R}\wedge\text{tr}\,\mathcal{R}\Big]\;.
\end{split}
\end{equation}
Having proved that $\sum_{m=0}^d(-)^mt_{n,m}(\beta)=\sum_{m=0}^d(-)^mh^{n,m}(\mathcal{M})$, it is evident that summing further on $n$ with an extra factor of $(-)^n$ provides the alternate sum of the real Betti numbers, yielding the Euler characteristics
\begin{equation}\label{t as euler}
\sum_{n,m=0}^d(-)^{m+n}t_{n,m}(\beta)=\sum_{n,m=0}^d(-)^{m+n}h^{n,m}(\mathcal{M})=\sum_{k=0}^{2d}(-)^kb_k(\mathcal{M})=\text{ind}(d)=\chi(\mathcal{M})\;,
\end{equation}
which is given as the integral of the Euler class $e(\mathcal{M})$, defined as the Pfaffian of the riemannian curvature
\begin{equation}
\chi(\mathcal{M})=\int_{\mathcal{M}}e(\mathcal{M})=\int_{\mathcal{M}}\text{Pf}\left(\frac{\mathcal{R}^{(G)}}{2\pi}\right)\;.
\end{equation}
Again we can check our formulas in $d=2$, where it gives
\begin{equation}\label{euler d=2}
\begin{split}
&\chi(\mathcal{M})=\frac{1}{32\pi^2}\int_{\mathcal{M}}\Big[\epsilon^{ABCD}\mathcal{R}^{(G)}_{AB}\wedge\mathcal{R}^{(G)}_{CD}\Big]\;,
\end{split}
\end{equation}
with $A,B,..$ being flat $SO(2d)$ indices.

There is still another useful identity that we need in section \ref{sec dualities} to prove the mismatch formula \eqref{mismatch}. Let us consider the sum
$\sum_{n,m=0}^d(-)^{n+m}n\,t_{n,m}(\beta)$. Using the symmetry under Hodge duality of the $t_{n,m}$ coefficients, and renaming variables, one gets
\begin{equation}
\begin{split}
\sum_{n,m=0}^d(-)^{n+m}n\,t_{n,m}(\beta)&=\sum_{n,m=0}^d(-)^{n+m}n\,t_{d-n,d-m}(\beta)=\sum_{n,m=0}^d(-)^{n+m}(d-n)t_{n,m}(\beta)\\
&=d\,\chi(\mathcal{M})-\sum_{n,m=0}^d(-)^{n+m}n\,t_{n,m}(\beta)
\end{split}
\end{equation}
proving the identity
\begin{equation}\label{t as second euler}
\sum_{n,m=0}^d(-)^{n+m}n\,t_{n,m}(\beta)=\sum_{n=0}^d(-)^nn\,\text{ind}\left(\Omega^{n,0},\bar\de\right)=\frac{d}{2}\,\chi(\mathcal{M})\;.
\end{equation}

As a final discussion on topological quantities, we investigate now the identification between non-propagating top forms and the analytic Ray-Singer torsion.
The effective action for a $(p,d-1)$-form can be written from \eqref{Z A} as
\begin{equation}\label{top form manipulated}
\begin{split}
Z^{A\,\text{top}}_{p,d-1}&=\int_0^\infty\frac{d\beta}{\beta}\sum_{n=0}^p\sum_{m=0}^{d-1}(-)^{n+m}(n+1)(m+1)t_{p-n,d-1-m}\\
&=\int_0^\infty\frac{d\beta}{\beta}\sum_{n=0}^p\sum_{m=0}^{d}(-)^{n+m+1}(n+1)m\,t_{d-p+n,m}\;.
\end{split}
\end{equation}
We focus now on the part
$$
\int_0^\infty\frac{d\beta}{\beta}\sum_{m=0}^{d}(-)^{m}m\,t_{d-p+n,m}\;.
$$
In order to define the torsion \cite{Ray:1973sb}, we recall that our hamiltonian coincides, up to a sign, with the Dolbeault laplacian and we have $H=(\bar\de\bar\de^\dagger+\bar\de^\dagger\bar\de)$. We shall denote with $\lambda_n$ the eigenvalues of $H$, that are non-negative, and define the corresponding zeta function as
\begin{equation}
\zeta_{p,q}(s)=\sum_{\lambda_n>0}\lambda_n^{-s}\;,
\end{equation}
where the subscript reminds that the hamiltonian is acting on $(p,q)$-forms.
By taking the derivative with respect to $s$ one can perform several manipulations
\begin{equation}
\zeta'_{p,q}(0)=-\sum_{\lambda_n>0}\ln(\lambda_n)=-\Tr'_{\mathcal{H}_{p,q}}\ln\,H=\int_0^\infty\frac{d\beta}{\beta}\Tr'_{\mathcal{H}_{p,q}}
\Big[{\rm e}^{-\beta H}\Big]=\int_0^\infty\frac{d\beta}{\beta}t_{p,q}(\beta)\;.
\end{equation}
In the last step we rewrote the logarithm as a Schwinger proper time integral. The prime over the trace dictates to subtract zero mode contributions,
that are present only if one considers a compact manifold.
Viewing the formula as a field theory effective action, this
subtraction corresponds to regulating the infrared divergences due to massless fields  and related to the large $\beta$ region.
 At this stage we can introduce the Ray-Singer analytic $\bar\de$-torsion
\begin{equation}\label{torsion}
\ln\,T_p(\mathcal{M})=\frac12\sum_{q=0}^d(-)^qq\,\zeta'_{p,q}(0)\;,
\end{equation}
and we readily recognize its relation with top forms
\begin{equation}\label{top form as torsion}
Z^{A\,\text{top}}_{p,d-1}=2\sum_{n=0}^p(-)^{n+1}(n+1)\ln T_{d-p+n}(\mathcal{M})\;.
\end{equation}
An analogous identification arises for ordinary differential forms \cite{Schwarz:1984wk}.

\end{document}